\documentclass[twocolumn,aps,amsmath,amssymb,floatfix,amsmath,superscriptaddress]{revtex4-1}

\usepackage{color,epsfig}
\usepackage{bm}
\usepackage{amsmath,amsfonts,amssymb,bm}
\usepackage{hyperref}
\hypersetup{
    colorlinks,%
    citecolor=blue,%
    linkcolor=blue,%
    urlcolor=blue
}

\usepackage{subfigure}
	\usepackage{feynmf}
\usepackage{calrsfs}
\usepackage{natmove}  %% automatically fixes punctuation around cites
\usepackage{dcolumn}% Align table columns on decimal point
\usepackage{multirow}
\usepackage{sidecap}
\usepackage{xcolor,cancel}
\usepackage[normalem]{ulem}
\usepackage{array}
\usepackage{float}
\usepackage{subfigure}
\usepackage{dsfont}
\usepackage{txfonts}
\usepackage{wasysym}
\DeclareMathOperator{\Tr}{Tr}

\newcommand{\bea}{\begin{eqnarray}}
\newcommand{\eea}{\end{eqnarray}}
\newcommand{\be}{\begin{equation}}
\newcommand{\ee}{\end{equation}}

\renewcommand\Re{\text{Re}}

\begin{document}
	
	\title{Electronic transport in Weyl semimetals with a uniform concentration of torsional dislocations}
	
	\author{Daniel A. Bonilla}
	\email{dabonilla@uc.cl}
	\affiliation{Facultad de F\'isica, Pontificia Universidad Cat\'olica de Chile, Vicu\~{n}a Mackenna 4860, Santiago, Chile}
	
	\author{Enrique Mu\~{n}oz}
	\email{ejmunozt@uc.cl}
	\affiliation{Facultad de F\'isica, Pontificia Universidad Cat\'olica de Chile, Vicu\~{n}a Mackenna 4860, Santiago, Chile}

	%%%%%%%%%%%%%%%%%%%%%%%%%%%%%%%%%%%%%%%%%%%%%%%%%%%%%%%%%%%%%%%%%%%%%%%%%%%%%%%%%%%%
%-------------------------	
	\begin{abstract}
	    In this article, we consider a theoretical model for a type I Weyl semimetal, under the presence of a diluted uniform concentration of torsional dislocations. By a mathematical analysis for partial wave scattering (phase-shift) for the T-matrix, we obtain the corresponding retarded and advanced Green's functions that include the effects of multiple scattering events with the ensemble of randomly distributed dislocations. Combining this analysis with the Kubo formalism, and including vertex corrections, we calculate the electronic conductivity as a function of temperature and concentration of dislocations. We further evaluate our analytical formulas to predict the electrical conductivity of several transition metal monopnictides, i.e. TaAs, TaP, NbAs and NbP.
	\end{abstract}
%---------------------------------	
	\maketitle
	
%-------------------------	
\section{Introduction}
\label{sec:introduction}
Weyl semimetals (WSMs) constitute a remarkable example of three-dimensional, gapless materials with nontrivial topological properties. First proposed theoretically, \cite{Xiangang2011,Fang2012,Ruan2016,Dirac,Felser,3dWeyl,Burkov} and, more recently, discovered experimentally on TaAs crystals\cite{Xu613}.

In a WSM, the band structure possesses an even number of Weyl nodes with linear dispersion, where
the conduction and valence bands touch. These nodes are monopolar sources of Berry curvature, and hence are protected from being gapped since their charge (chirality) is a topological invariant \cite{Burkov}. In the vicinity of these nodes, low energy conducting states behave as Weyl fermions, i.e. massless quasi-particles with pseudo-relativistic Dirac linear dispersion \cite{Dirac,Felser,3dWeyl,3dWeyl,Burkov}. In Weyl fermions, conserved chirality determines the projection of spin over their momentum direction, a condition referred to as ``spin-momentum locking''. While Type I WSMs are Lorentz covariant, this symmetry is violated in Type II WSMs, where the Dirac cones are strongly tilted \cite{vanderbilt}.\\

The presence of Weyl nodes in the bulk spectrum determines the emergence of Fermi arcs\cite{Xu613}, the chiral anomaly, and the chiral magnetic effect, among other remarkable properties \cite{vanderbilt}.  Therefore, considerable attention has been paid to understand the electronic transport properties of WSMs \cite{Hosur2013,Hu2019,Nagaosa2020}. For instance, there are recent works on charge transport \cite{Hosur2012} in the presence of spin-orbit coupled impurities \cite{Liu2017}, electrochemical\cite{Flores2021} and nonlinear transport induced by Berry curvature dipoles \cite{Zeng2021}.  Somewhat less explored are the effects of mechanical strain and deformations in WSMs. From a theoretical perspective, it has been proposed that different types of elastic strains can be modeled as gauge fields in WSMs \cite{Cortijo_2015,Cortijo_2016,Arjona_Vozmediano_PRB2018}. In previous works, we have studied the combined effects of a single torsional dislocation and an external magnetic field on the electronic \cite{Munoz2019,Soto_Garrido_2020} and thermoelectric \cite{Munoz2019,Bonilla_Munoz_2021} transport properties of WSMs, using the Landauer ballistic formalism in combination with a mathematical analysis for the quantum mechanical scattering cross-sections\cite{Munoz_2017}.\\

In this work, we extend our previous analysis to study the case of a diluted, uniform concentration of torsional dislocations and its effects on the electrical conductivity of type I WSMs. In contrast to our former studies\cite{Bonilla_Munoz_2021,Munoz2019,Soto_Garrido_2020}, here we employ the Kubo linear-response formalism at finite temperatures, that therefore requires to explicitly calculate the retarded and advanced Green´s functions for the system, including the multiple scattering events due to the random distribution of dislocation defects in the form of a disorder-averaged self-energy term. For this purpose, we first analyze the phase shift arising from a single torsional dislocation, and obtain the corresponding (retarded and advanced) Green's function in terms of the T-matrix elements by solving analytically the Lippmann-Schwinger equation. We further extend this analysis, incorporating the effect of a random distribution of such dislocations, with a concentration $n_d$, in the form of a disorder-averaged self-energy into the corresponding Dyson's equation. Finally, we analyze the correction to the scattering vertex, and by including this additional contribution we calculate the electrical conductivity from the Kubo formula, as a function of temperature and concentration of dislocations. We present explicit evaluations of our analytical expressions for the electrical conductivity as a function of temperature and concentration of dislocations $n_d$, for several materials in the family of transition metals monopnictides, i.e. TaAs, TaP, NbAs and NbP, where the corresponding microscopic parameters, estimated by ab-initio methods, where reported in the literature\cite{Szot_2018,Lee_2015,Grassano_2018}.

%---------------------------
\section{Scattering by a single dislocation}
\begin{figure}[ht]
    	\centering
    	\includegraphics[width= \columnwidth]{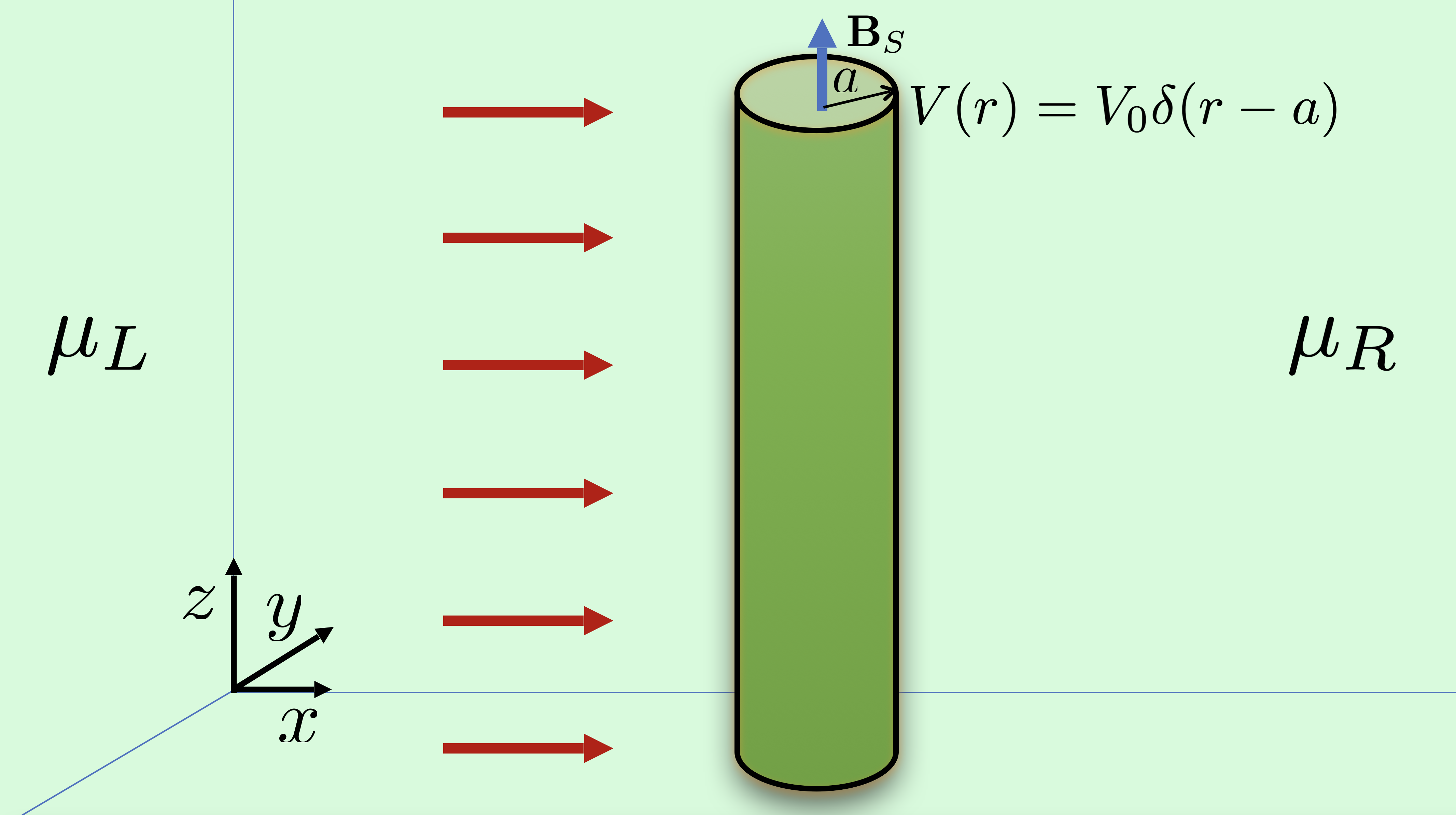}
    	\caption{Pictorial description of the scattering of free incident Weyl fermions coming from a left reservoir by a single cylindrical dislocation defect. }
    	\label{fig:single_defect}
    \end{figure}

As a continuum model for a type I WSM under the presence of a single dislocation defect, as depicted in Fig.~\eqref{fig:single_defect}, we consider the Hamiltonian\cite{Bonilla_Munoz_2021}
\begin{equation}
		\hat{H}^{\xi}=\xi\hbar v_F \boldsymbol{\sigma}\cdot\left( \mathbf{p} + e\mathbf{A}^{\xi}
			\right) + \sigma_0 V_0\delta(r-a)
			\equiv \hat{H}^{\xi}_0+\hat{H}^{\xi}_1, \label{eq:hamiltonian}
\end{equation}
where
\begin{eqnarray}
		\hat{H}^{\xi}_0&=&\xi v_F \boldsymbol{\sigma}\cdot\mathbf{p}, \label{eq:H_0} \\
		\hat{H}^{\xi}_1&=& \xi e v_F \left( \boldsymbol{\sigma}\cdot\boldsymbol{\hat{\phi}} \right) \frac{1}{2}B^{\xi}r\Theta(a-r)+V_0\delta(r-a)\sigma_0. \label{eq:H_1}
\end{eqnarray}
Here, $\xi=\pm$ labels each of the Weyl nodes located at $\mathbf{K}_{\pm}=\pm \mathbf{b}/2$. The expression in Eq.~\eqref{eq:H_0} is the free-particle Hamiltonian,  whereas the expression in Eq.~\eqref{eq:H_1} represents the interaction with the dislocation, where torsional strain is described as a pseudo-magnetic field inside the cylinder\cite{Munoz_2017,Bonilla_Munoz_2021,Soto_Garrido_2020}, as well as the lattice mismatch effect at the boundary of the dislocation, modeled as
repulsive delta barrier on its surface\cite{Bonilla_Munoz_2021}.
%---------------------------
%------------------------------
The "free" spinor eigenfunctions for the defect-free reference system satisfy
\begin{equation}
		\hat{H}_{0}^{\xi} \left|\Phi_{\mathbf{k},\lambda} \right\rangle  =\mathcal{E}^{(0,\xi)}_{\lambda,\mathbf{k}} \left|\Phi_{\mathbf{k},\lambda}\right\rangle, \label{eq:eigen_eq_dirac}
	\end{equation}
where the energy spectrum is given by
	\begin{equation}
		\mathcal{E}^{(0,\xi)}_{\lambda,\mathbf{k}}=\lambda\xi\hbar v_F |\mathbf{k}|, \label{eq:unpert_spectrum}
	\end{equation}
and $\lambda=\pm 1$ is the band (helicity) index. When projected onto coordinate space, these spinor eigenfunctions have the explicit form
\begin{equation}
		\Phi_{\lambda,\mathbf{k}}(\mathbf{r})=\sqrt{\frac{|\mathbf{k}|+\lambda k_z}{2|\mathbf{k}|}}\begin{pmatrix}
			1 \\ \frac{\lambda|\mathbf{k}|-k_z}{k_x-ik_y}
		\end{pmatrix}e^{i\mathbf{k}\cdot \mathbf{r}},\label{eq:spinor_final}
	\end{equation}
and constitute an orthonormal basis for the Hilbert space.

If we now consider the (elastic) scattering effects induced by the torsional dislocation modeled by Eq.~\eqref{eq:H_1}, we need to look for the eigenvectors
$\left|\Psi_{\lambda,\mathbf{k}}\right\rangle$ of the total Hamiltonian in Eq.~\eqref{eq:hamiltonian} with the same energy as in Eq.~\eqref{eq:unpert_spectrum}. The answer is provided by the solution to the well known Lippmann-Schwinger equation
		\begin{equation}
			\left| \Psi_{\mathbf{k},\lambda}\right\rangle = \left|\Phi_{\mathbf{k},\lambda}\right\rangle  + \hat{G}^{\xi}_{R,0}(E)\hat{H}_{1}^{\xi} \left| \Psi_{\mathbf{k},\lambda}\right\rangle, \label{eq:Lippmann_Schwinger}
		\end{equation}
where the free Green's function can be expressed in a coordinate-independent representation form via the \textit{resolvent}, 
	\begin{equation}
		\hat{G}^{\xi}_{R/A,0}(E)=\left[E-\hat{H}^{\xi}_0\pm i\eta^{+}\right]^{-1}. \label{eq:resolvent}
	\end{equation}
Here, the index $R/A$ stands for retarded and advanced, respectively. As shown in detail in the Appendix \ref{app:GF}, in the coordinate representation the corresponding free Green's function is given by the explicit matrix form $\mathbf{G}^{\xi }_{R,0}\left( \mathbf{r},\mathbf{r'};k\right)=\delta(z-z')\mathbf{G}^{\xi }_{R,0}\left( \mathbf{x},\mathbf{x'};k\right)$, where $\mathbf{r} = (\mathbf{x},z)$ and
\begin{align}
	\mathbf{G}^{\xi }_{R,0}\left( \mathbf{x},\mathbf{x'};k\right)&= -\frac{\lambda \xi ik}{4 \hbar v_F}\nonumber \\
  &\times \begin{bmatrix}
			H^{(1)}_0 \left(k|\mathbf{x}-\mathbf{x'}| \right)& i\lambda e^{-i\varphi} H^{(1)}_1 \left(k|\mathbf{x}-\mathbf{x'}|\right) \\
			i\lambda e^{i\varphi} H^{(1)}_1 \left(k|\mathbf{x}-\mathbf{x'}|\right)& H^{(1)}_0 \left(k|\mathbf{x}-\mathbf{x'}| \right)
		\end{bmatrix}. \label{eq:G0_final}
	\end{align}
Here, $H_{0}^{(1)}(z)$ and $H_{1}^{(1)}(z)$ are the Hankel functions and $\mathbf{x}=(x,y)$ is the position vector on any plane perpendicular to the cylinder's axis.

For the scattering analysis, we need the retarded resolvent for the full Hamiltonian, which is defined as the solution to the equation
        \begin{equation}
			\left( E + i\eta^{+}-\hat{H}^{\xi}\right) \hat{G}^{\xi}_{R}(E)=\hat{I}.
			\label{eq:GR}
		\end{equation}
%------------------------------------
	\begin{figure}[ht]
		\centering
		\includegraphics[width= \columnwidth]{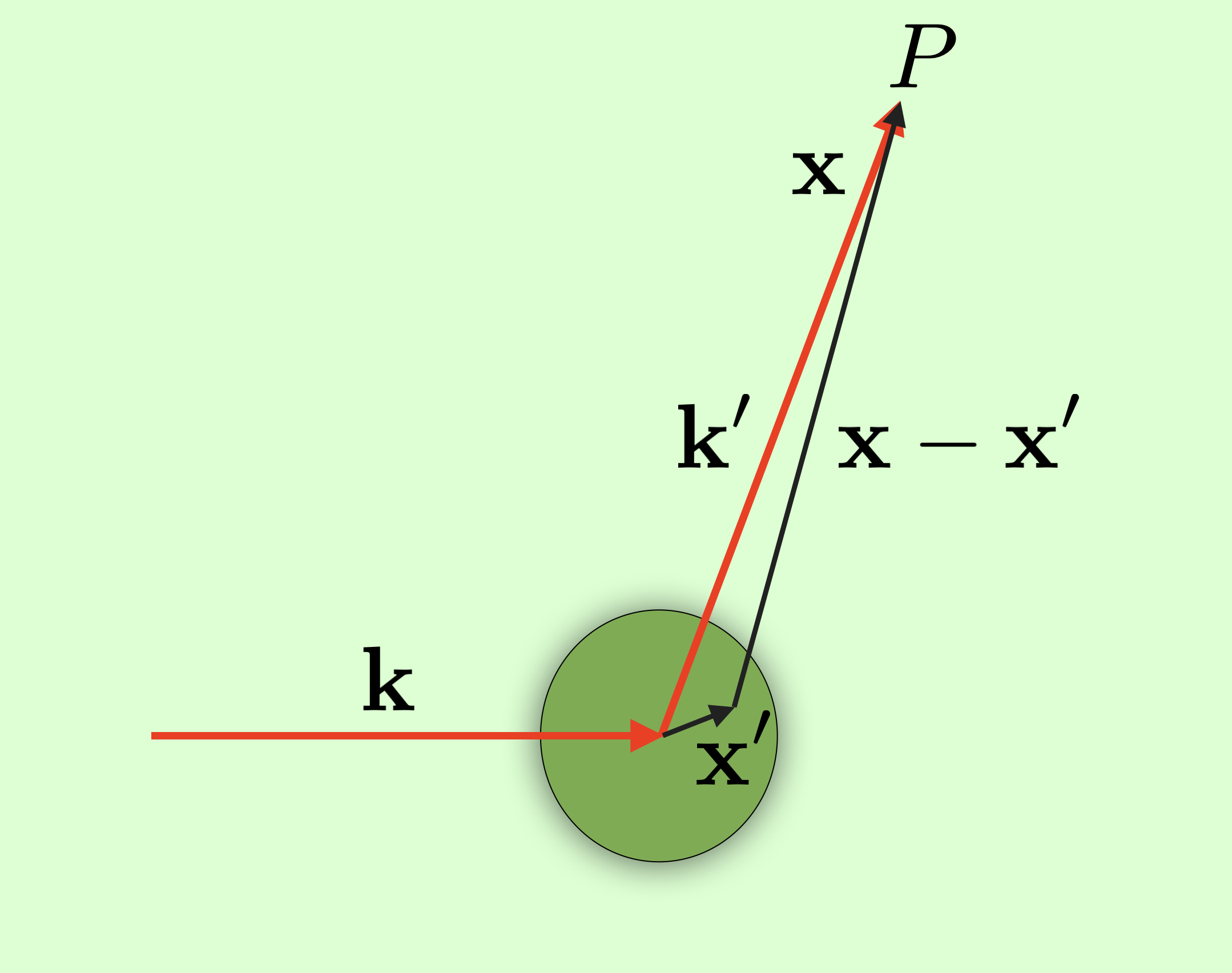}
		\caption{Pictorial description of the scattering event on a plane perpendicular to the cylindrical defect axis.}
		\label{fig:scattering}
	\end{figure}
%----------------------------------
Combining Eq.~\eqref{eq:GR} with Eq.~\eqref{eq:resolvent}, we readily obtain
	\begin{eqnarray}
		\hat{G}^{\xi}_{R}(E)&=&\hat{G}^{\xi}_{R,0}(E)+\hat{G}^{\xi}_{R,0}(E)\hat{H}^{\xi}_1\hat{G}^{\xi}_{R}(E)\nonumber\\
		&=&\hat{G}^{\xi}_{R,0}(E)+\hat{G}^{\xi}_{R,0}(E)\hat{T}^{\xi}(E)\hat{G}^{\xi}_{R,0}(E),
	\end{eqnarray}
where we introduced the standard definition of the T-matrix operator $\hat{T}^{\xi}(E)$, that can be formally expressed in closed form by
\begin{eqnarray}
	  \hat{T}^{\xi}(E) &=&  \hat{H}_1^{\xi}+\hat{H}_1^{\xi}\hat{G}^{\xi}_{R,0}(E) \hat{T}^{\xi}(E)\nonumber\\
	  &=& \hat{H}_{1}^{\xi}\left(\hat{I}- \hat{G}^{\xi}_{R,0}(E)\hat{H}_{1}^{\xi}\right) ^{-1}.
\end{eqnarray}
Using this definition, along with the property $\hat{H}_{1}^{\xi}|\Psi_{\mathbf{k},\lambda}\rangle = \hat{T}^{\xi}|\Phi_{\mathbf{k},\lambda} \rangle$, 
we obtain the Lippmann-Schwinger Eq.~\eqref{eq:Lippmann_Schwinger} in the coordinate representation
		\begin{eqnarray}
		\Psi_{\mathbf{k},\lambda}(\mathbf{r}) = \Phi_{\mathbf{k},\lambda}(\mathbf{r})\,   +&&\int d^3r' \int d^3r'' \left\langle\mathbf{r}\middle|\hat{G}^{\xi}_{R,0}\left( E \right)\middle|\mathbf{r'}\right\rangle\nonumber\\
				&&\times\left\langle\mathbf{r'}\middle|\hat{T}^{\xi}(E)\middle|\mathbf{r''}\right\rangle \Phi_{\mathbf{k},\lambda}(\mathbf{r''}). \label{eq:Lippmann_Schwinger_final}
	\end{eqnarray}
As shown in detail in Appendix \ref{app:scattering}, by considering the asymptotic behavior of the Hankel functions, $H^{(1)}_{\nu}(x)\sim \sqrt{\frac{2}{\pi x}}e^{i\left(x-\frac{\nu \pi}{2}-\frac{\pi}{4} \right) }$ (for $x\rightarrow \infty.$), Eq.~\eqref{eq:Lippmann_Schwinger_final} can be reduced to the $x$-$y$ plane and takes the explicit asymptotic expression
	\begin{equation}
		\Psi_{\mathbf{k}_{\parallel},\lambda}(\mathbf{x})\sim\frac{1}{\sqrt{2}} \begin{pmatrix}
			1 \\ \lambda 
		\end{pmatrix}e^{ikx}-\frac{\lambda \xi }{2 \hbar v_F}\sqrt{\frac{i k}{\pi}}T^{(\lambda,\xi)}_{\mathbf{k'}_{\parallel}\mathbf{k}_{\parallel}} \begin{pmatrix}
			1 \\ \lambda e^{i\phi}
		\end{pmatrix}\frac{e^{ikr}}{\sqrt{r}}, \label{eq:assymptotic_wave_fn}
	\end{equation}
where as we explain in the Appendix, the particles have only momenta perpendicular to the defect's axis, i.e., $\mathbf{k}_{\parallel}=(k_x,k_y)$. Comparing this last result with our previous reported expression for the scattering amplitude \cite{Soto_Garrido_2018}
	\begin{equation}
		\begin{bmatrix}
			f_{1}(\phi)  \\ f_{2}(\phi)
		\end{bmatrix}=\frac{e^{-\frac{i\pi}{4}}}{\sqrt{4 \pi k}}\sum_{m=-\infty}^{\infty} \begin{bmatrix}
			e^{i m \phi} \\ \lambda e^{i(m+1)\phi}
		\end{bmatrix}\left(e^{2i\delta_m}-1 \right), \label{eq:scatt_amplitudes}
	\end{equation}
we identify $T^{(\lambda,\xi)}_{\mathbf{k'}\mathbf{k}}=-2 \lambda \xi \hbar v_F \sqrt{\pi/i k} f_1(\phi)$. Therefore, we arrived at an explicit analytical expression for the T-matrix elements in terms of the phase shift $\delta_{m}(k)$ for each angular momentum channel $m\in\mathbb{Z}$
	\begin{equation}
		T^{(\lambda,\xi)}_{\mathbf{k'}_{\parallel}\mathbf{k}_{\parallel}}=-\frac{2 \lambda \xi  \hbar v_F}{k}\sum_{m=-\infty}^{\infty}e^{i\delta_m(k)}\sin \delta_m(k)e^{im\phi},\label{eq:Tmatrix_final}
	\end{equation}
	where $\phi$ is the angle between $\mathbf{k}_{\parallel}$ and $\mathbf{k'}_{\parallel}$, and the analytical expression for the phase shift is given in Appendix \ref{app:scattering} by Eq.~\eqref{eq:phase_shift}.
%-------------------------------	
\section{Scattering by a uniform concentration of dislocations}
%-----------------------------------
\begin{figure}[ht]
		\centering
		\includegraphics[width=\columnwidth]{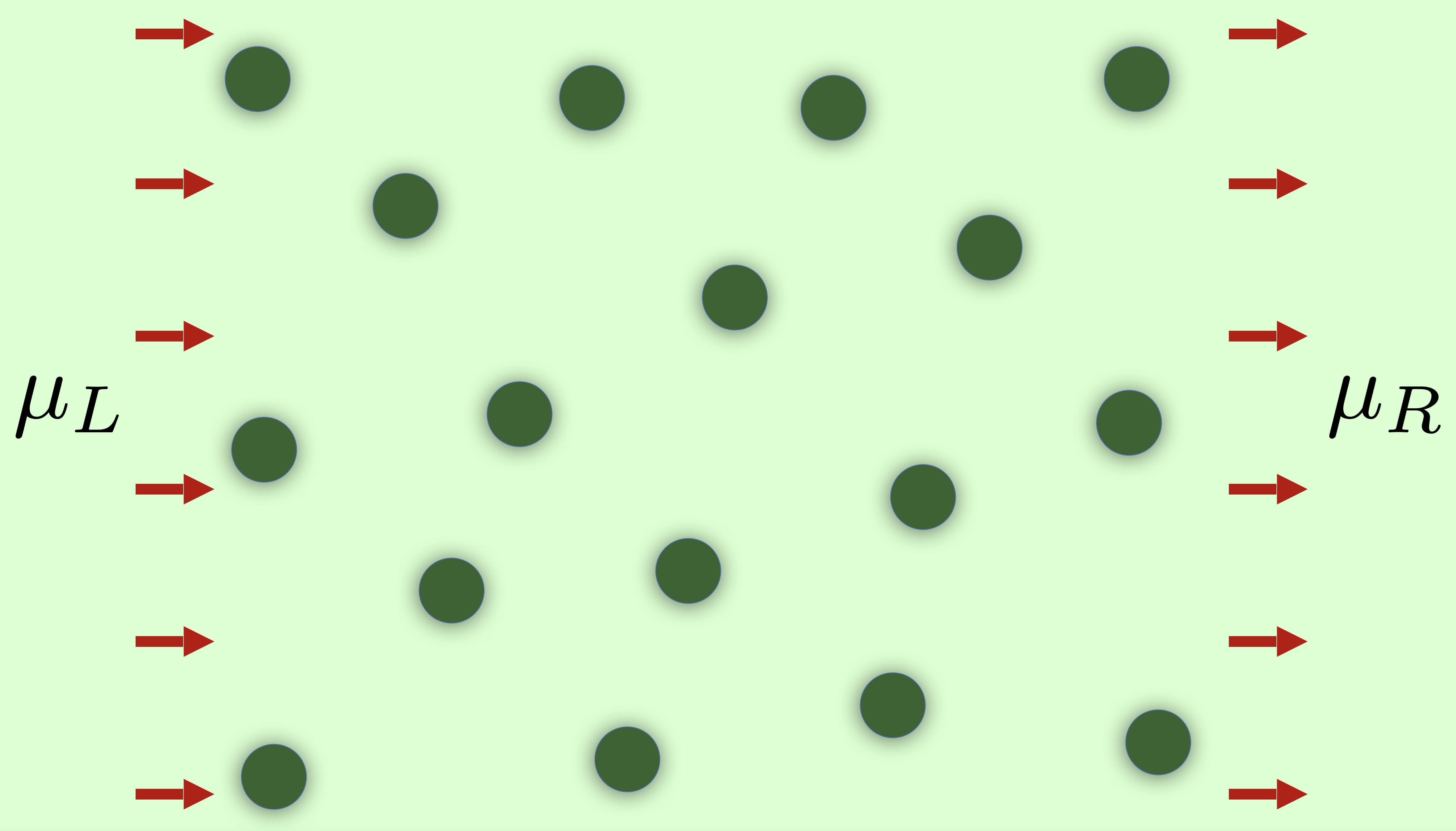}
		\caption{Random distribution of torsional dislocations seen from a plane perpendicular to the cylinders axis.}
		\label{fig:several_dislocations}
	\end{figure}
%-----------------------------------
Let us now consider a uniform concentration $n_d = N_d/A$ (per unit transverse surface) of identical cylindrical dislocations, as depicted in FIG.~\ref{fig:several_dislocations}, represented by the density function
\begin{equation}
\rho(\mathbf{x})=\sum_{j=1}^{N_d}\delta(\mathbf{x} - \mathbf{X}_j),
\end{equation}
where $\mathbf{X}_j$ is the position of the $j$-defect's axis. The Fourier transform of this density function is thus given by the expression
\begin{equation}
	    \tilde{\rho}(\mathbf{k}_{\parallel})= \int d^2 x e^{-i \mathbf{k}_{\parallel}\cdot\mathbf{x}} \rho(\mathbf{x}) =\sum_{j=1}^{N_d} e^{-i\mathbf{k}_{\parallel}\cdot \mathbf{X}_j}.\label{eq:rho_def}
\end{equation}
The operator that plays the role of a scattering potential for this distribution of dislocation defects is
	\begin{equation}
		V(\mathbf{x})= \int d^2 x' \rho(\mathbf{x'}) H^{\xi}_{1}(\mathbf{x}-\mathbf{x}') = 
		\sum_{j=1}^{N_d}H^{\xi}_{1}(\mathbf{x}-\mathbf{X}_j), \label{eq:Potential_several_defects}
	\end{equation}
	where $H^{\xi}_1$ is defined in Eq.~\eqref{eq:H_1} as the contribution from a single dislocation. The matrix elements of the scattering operator Eq.~\eqref{eq:Potential_several_defects} in the free spinor basis defined by Eq.~\eqref{eq:eigen_eq_dirac} are
	\begin{align}
	\left\langle \Phi_{\mathbf{k}_{\parallel},\lambda}
	\middle|V(\mathbf{x})\middle|\Phi_{\mathbf{k'}_{\parallel},\lambda'}\right\rangle=\left[\tilde{V}(\mathbf{k}_{\parallel}-\mathbf{k'}_{\parallel})\right]_{\lambda\lambda'},\label{eq:matrix_element_V}
	\end{align}
	where $\tilde{V}(\mathbf{k}_{\parallel})$ is the Fourier transform
	\begin{align}
		\tilde{V}(\mathbf{k}_{\parallel})&=\int_{\mathbb{R}^2} d^2 x e^{-i\mathbf{k}_{\parallel}\cdot \mathbf{x}}V(\mathbf{x})=\sum_{j=1}^{N_d}\int_{\mathbb{R}^2} d^2	 x e^{-i\mathbf{k}_{\parallel}\cdot \mathbf{x}}H^{\xi}_{1}(\mathbf{x}-\mathbf{X}_j)\nonumber\\
		&=\tilde{H}^{\xi}_{1}(\mathbf{k}_{\parallel})\tilde{\rho}(\mathbf{k}_{\parallel}).\label{eq:fourier_V}
	\end{align}
	Then, the matrix elements of the potential in Eq.~\eqref{eq:matrix_element_V} become 
	\begin{equation}
	    \left[\tilde{V}(\mathbf{k}_{\parallel})\right]_{\lambda\lambda'}= \left[\tilde{H}^{\xi}_{1}(\mathbf{k}_{\parallel})\right]_{\lambda\lambda'}\tilde{\rho}(\mathbf{k}_{\parallel}). \label{eq:potential_disorder}
	\end{equation}
	Let us also introduce the configurational average of some quantity $f(\mathbf{X}_j)$ over the distributed dislocations as
	\begin{equation}
		\left\langle f \right\rangle=\int_{\mathbb{R}^2}d^2X_j \,\,P(\mathbf{X}_j)f(\mathbf{X}_j), \label{eq:conf_average}
	\end{equation}
	where $P(\mathbf{X}_j)$ is the normalized distribution function for the defects in the sample. In particular,
	for a uniform distribution we have $P(\mathbf{X}_j)=1/A$,
	where $A$ is the area of the plane normal to each cylinder's axis. 
	Now, the full retarded Green's function satisfies for the potential of several dislocations $\hat{V}$ given in Eq.~\eqref{eq:Potential_several_defects} is
	\begin{equation}
		\hat{G}^{\xi}_{R}(E)=\hat{G}^{\xi}_{R,0}(E)+\hat{G}^{\xi}_{R,0}(E)\hat{V}\hat{G}^{\xi}_{R}(E).
	\end{equation}
	The configurational average, as defined in Eq.~\eqref{eq:conf_average}, of the complete Green's function in this last equation can be written as 
   \begin{equation}
   	\left\langle\hat{G}^{\xi}_{R}(E)\right\rangle=\hat{G}^{\xi}_{R,0}(E)+\hat{G}^{\xi}_{R,0}(E)\Sigma_{R}^{\lambda,\xi}(E)\left\langle\hat{G}^{\xi}_{R}(E)\right\rangle. \label{eq:Dyson_eqn}
   \end{equation}
   This is the Dyson's equation with the retarded self-energy $\Sigma_{R}^{\lambda,\xi}(E)$, that can be explicitly solved to yield
   \begin{eqnarray}
       \left\langle G_R^{\lambda,\xi}(\mathbf{k}_{\parallel})\right\rangle=\frac{1}{ E-\lambda\xi\hbar v_F |\mathbf{k}_{\parallel}|-\Sigma_R^{\lambda,\xi}(\mathbf{k}_{\parallel})}. \label{eq:GR_final}
   \end{eqnarray}
	
\begin{figure}[!ht]
		\centering
		\includegraphics[width=0.8 \columnwidth]{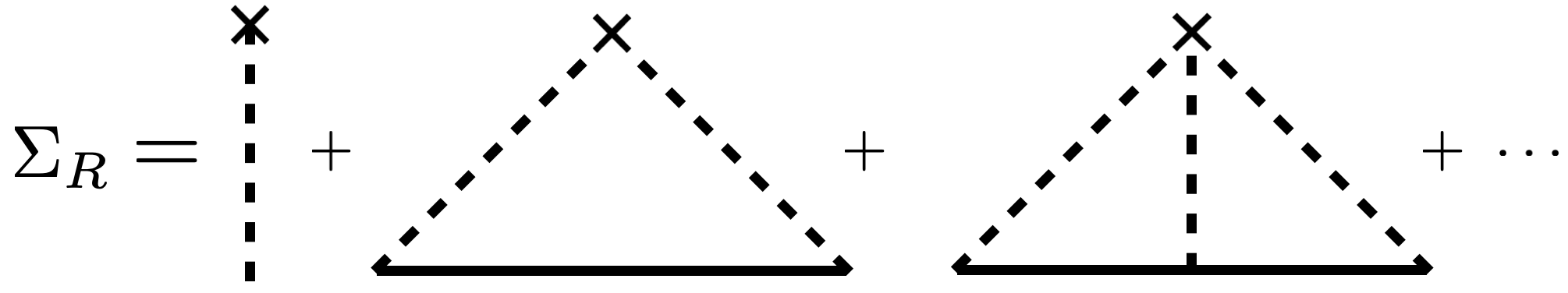}
		\caption{Diagrams contributing to the retarded self-energy $\Sigma_R$. The solid line corresponds to the free retarded Green's function, the dashed line the scattering perturbation $H_1$, and the $\times$ a factor of $n_d$.}
		\label{fig:diagrams}
	\end{figure}

	The effect of the statistical distribution of dislocations' is entirely dictated by the function $\tilde{\rho}(\mathbf{k}_{\parallel})$. In the perturbative expansion of the complete Green's function, we encounter $n^{th}$-products of the form $\tilde{\rho}(\mathbf{k_1})\tilde{\rho}(\mathbf{k_2})\cdots\tilde{\rho}(\mathbf{k}_{n})$. The configurational average of these products are
	\begin{eqnarray}
		\left\langle \tilde{\rho}(\mathbf{k}_{\parallel}) \right\rangle &=&\left\langle \sum_{j=1}^{N_d}e^{-i\mathbf{k}_{\parallel}\cdot \mathbf{X}_j} \right\rangle=\sum_{j=1}^{N_d}\int_{\mathbb{R}^2}d^2X_j \,\frac{1}{A}\,e^{-i\mathbf{k}_{\parallel}\cdot \mathbf{X}_j}\nonumber\\
		&=&\frac{N_d}{A}(2\pi)^2\delta^{(2)}(\mathbf{k}_{\parallel}),\label{eq:rho_1}
	\end{eqnarray}
	for a single factor. For the product of two factors, we obtain
	\begin{eqnarray}
		&&\left\langle \tilde{\rho}(\mathbf{k}_{\parallel}^1) \tilde{\rho}(\mathbf{k}_{\parallel}^2) \right\rangle=\left\langle \sum_{j=1}^{N_d}\sum_{l=1}^{N_d}e^{-i\mathbf{k}_{\parallel}^1\cdot \mathbf{X}_j-i \mathbf{k}_{\parallel}^2\cdot \mathbf{X}_l}\right\rangle\\
		&=&\left\langle \sum_{j=l}e^{-i(\mathbf{k}_{\parallel}^1+\mathbf{k}_{\parallel}^2)\cdot \mathbf{X}_j}+\sum_{j\neq l}e^{-i\mathbf{k}_{\parallel}^1\cdot \mathbf{X}_j-i \mathbf{k}_{\parallel}^2\cdot \mathbf{X}_l}\right\rangle\nonumber\\
		&=&\frac{N_d}{A}(2\pi)^2\delta^{(2)}(\mathbf{k}_{\parallel}^1+\mathbf{k}_{\parallel}^2)+ \frac{N_d(N_d-1)}{A^2}(2\pi)^4 \delta^{(2)}(\mathbf{k}_{\parallel}^1)\delta^{(2)}(\mathbf{k}_{\parallel}^2),\nonumber\label{eq:rho_2}
	\end{eqnarray}
	and we have a similar behavior for higher order products. Now, notice that for $N_d\gg 1$  we have $N_d(N_d-1)\approx N_d^2$, $N_d(N_d-1)(N_d-3)\approx N_d^3$ and so on. We define the concentration of defects, i.e., the number of dislocations per unit of area perpendicular to the cylinder's axis as $n_d=N_d/A$. As discussed in standard references \cite{Hewson,Mahan}, for small concentrations $n_d \ll 1$ the scaling discussed before ensures that the total Green's function in Eq.~\eqref{eq:Dyson_eqn} can be calculated accurately by the sequence of diagrams for the retarded self-energy in momentum space as given in FIG.~\ref{fig:diagrams}, an approach well known as the non-crossing approximation (NCA). This series of diagrams corresponds to the configurational average of the $T$-matrix over the random distribution of dislocations after Eq.~\eqref{eq:conf_average}
   \begin{equation}
   	 \Sigma_{R}^{\lambda,\xi}(E)	=\left\langle\hat{T}^{\xi}(E)\right\rangle = n_d T^{(\lambda,\xi)}_{\mathbf{k}_{\parallel}\mathbf{k}_{\parallel}}.
   \end{equation}
Using the expression in Eq.~\eqref{eq:Tmatrix_final} for the $T$-matrix elements, for $\mathbf{k}_{\parallel}=\mathbf{k'}_{\parallel}$ then $\phi=0$ and we have that the real part of the self-energy
	\begin{equation}
	    \text{Re}\,\,\Sigma_R^{\lambda,\xi}(\mathbf{k}_{\parallel})=-\frac{2 \lambda \xi  n_d \hbar v_F}{k}\sum_{m=-\infty}^{\infty}\cos \delta^{(\lambda,\xi)}_m(k)\sin \delta^{(\lambda,\xi)}_m(k) \label{eq:real_part_self_energy}
	\end{equation}
	contains an infinite sum over highly oscillatory terms, that converges to zero. So no contribution comes from the real part of the self-energy; The imaginary part, on the other hand, defines the relaxation time,
	\begin{equation}
	    \frac{1}{\tau^{(\lambda,\xi)}(k)}=-\frac{2\lambda\xi}{\hbar}n_d\,\,\text{Im}\,\,T^{(\lambda,\xi)}_{\mathbf{k}_{\parallel}\mathbf{k}_{\parallel}}. \label{eq:tau}
	\end{equation}
\subsection{Electrical Conductivity in the linear-response}
	
	Now, we consider a single Fourier mode for an external electric field in the gauge
	\begin{equation}
		\mathbf{E}=-\frac{\partial}{\partial t}\mathbf{A}(\mathbf{r},t)
	\end{equation}
	where $\mathbf{A}(\mathbf{r},t)=\mathbf{A}(\mathbf{r},\omega)e^{-i\omega t}$ is the vector potential. Then, $\mathbf{E}=i\omega\mathbf{A}$.  In the linear response formalism, the current is given by the Kubo expression
	\begin{eqnarray}
	j_{\alpha}(\mathbf{r},\omega)	&=&\int d^3r' \,\sigma_{\alpha\beta}(\mathbf{r},\mathbf{r'};\omega)E_{\beta}(\mathbf{r'}.\omega), \label{eq:current_lrt}
	\end{eqnarray}
	Here, the conductivity tensor is given by
	\begin{equation}
		\sigma_{\alpha\beta}(\mathbf{r},\mathbf{r'};\omega)=\frac{1}{i\omega} K_{\alpha\beta}(\mathbf{r},\mathbf{r'};\omega).  \label{eq:def_conductivity}
	\end{equation}
	The tensor $K_{\alpha\beta}$ is defined, in the Kubo formalism, in terms of the retarded current-current correlator as follows
\begin{eqnarray}
K_{\alpha\beta}(\mathbf{r},t;\mathbf{r'},t')=i\hbar^{-1}\theta(t-t')\rm{Tr} \left\lbrace \hat{\rho}\left[ \hat{j}_{\alpha}(\mathbf{r},t),\hat{j}_{\beta}(\mathbf{r'},t')\right]\right\rbrace, \label{eq:corr_current}
\end{eqnarray}
where $\hat{\rho}$ is the statistical density matrix operator. As shown in detail in the Appendix \ref{app:linear}, the Fourier transform to the frequency domain of this tensor can be expressed by
\begin{eqnarray}
		K^{\xi}_{\alpha\beta}(\mathbf{r},\mathbf{r'};\omega)&=&e^2v_{F}^2\int_{-\infty}^{\infty} \frac{dE'}{2\pi}\int_{-\infty}^{\infty} \frac{dE}{2\pi} \frac{f_0 \left(E' \right)-f_0\left(E\right) }{\hbar\omega+ E-E'+i\eta^{+}}\nonumber\\
		&&\times\rm{Tr}\left[ \sigma_{\alpha}\boldsymbol{\mathcal{A}}^{\xi}(\mathbf{r},\mathbf{r'};E')\sigma_{\beta} \boldsymbol{\mathcal{A}}^{\xi}(\mathbf{r'},\mathbf{r};E)\right],\label{eq:corr_freq_spectral}
	\end{eqnarray}
where $f_{0}(E)=\left[e^{(E-\mu)/kT}+1\right]^{-1}$ is the Fermi distribution, and we introduced the (disorder-averaged) spectral function
\begin{eqnarray}
\mathcal{A}^{\lambda,\xi}(k)&=& i\left[ 	\left\langle G_R^{\lambda,\xi}(\mathbf{k}_{\parallel})\right\rangle-	\left\langle G_A^{\lambda,\xi}(\mathbf{k}_{\parallel})\right\rangle\right]\nonumber\\
&=&\frac{2\left(\frac{\hbar}{2\tau^{(\lambda,\xi)}(k)}  \right) }{\left(E-\mathcal{E}^{\lambda,\xi}_{k}\right) ^2+\left(\frac{\hbar}{2\tau^{(\lambda,\xi)}(k)} \right) ^2},
\end{eqnarray}
that clearly reduces to a Lorentzian distribution whose spectral width is defined by the inverse of the relaxation time. See Appendix \ref{app:spectral} for the details. After some algebraic manipulations, we obtain the conductivity tensor at finite frequency and temperature
\begin{eqnarray}
		&&\Re\,	\sigma^{\xi}_{\alpha\beta}(\mathbf{r},\mathbf{r'};\omega)
		=-\frac{e^2\hbar v_{F}^2}{2\pi} \int_{-\infty}^{\infty} dE 
		\left\lbrace \frac{f_{0}(E+\hbar\omega)-f_{0}(E)}{\hbar\omega}\right\rbrace\nonumber\\
		&&\times\rm{Tr}\left[ \sigma_{\alpha}\boldsymbol{\mathcal{A}}^{\xi}(\mathbf{r},\mathbf{r'};E+\hbar\omega)\sigma_{\beta}\boldsymbol{\mathcal{A}}^{\xi}(\mathbf{r'},\mathbf{r};E)\right]. 
	\end{eqnarray}
Using the coordinates representation of the spectral function given in Appendix \ref{app:spectral}, after Eq.~\eqref{eq:spectral_coordinates}, we can read off the Fourier transform to momentum space of the conductivity
\begin{align}
     &\sigma^{\xi}_{\alpha\beta}(\mathbf{q};\omega)=-\frac{e^2\hbar v_{F}^2}{2\pi }\int \frac{d^3k}{(2\pi)^3} \int_{-\infty}^{\infty} dE   \left\lbrace \frac{f_{0}(E+\hbar\omega)-f_{0}(E)}{\hbar\omega}\right\rbrace \nonumber\\
    &\quad \times \sum_{\lambda,\lambda'}\Tr \left\lbrace \sigma_{\alpha}\left( \sigma_0+\lambda\frac{\boldsymbol{\sigma\cdot}(\mathbf{k}_{\parallel}+\mathbf{q})}{|\mathbf{k}_{\parallel}+\mathbf{q}|}\right)\sigma_{\beta}\left( \sigma_0+\lambda'\frac{\boldsymbol{\sigma\cdot}\mathbf{k}_{\parallel}}{|\mathbf{k}_{\parallel}|}\right) \right\rbrace\nonumber\\
    &\quad \times \mathcal{A}^{\lambda,\xi}(|\mathbf{k}_{\parallel}+\mathbf{q}|;E+\hbar\omega)\mathcal{A}^{\lambda',\xi}(|\mathbf{k}_{\parallel}|;E). \label{eq:condcutivity_q}
 \end{align} 
We are interested in the DC conductivity, so we take the limit $\mathbf{q}\rightarrow\mathbf{0}$ first and then the limit $\omega\rightarrow0$. After a long calculation (details in the Appendix \ref{app:linear}), the result is
\begin{align}
	    \sigma_{\alpha\beta}^{(\lambda,\xi)}(T)&= \delta_{\alpha\beta}\frac{e^2\hbar v_{F}^2}{\pi^3 }\int_{0}^{\infty} dk \,\, \int_{-\infty}^{\infty}dE\,\,\left(-\frac{\partial f_0(E)}{\partial E}\right) \notag\\
    &\times \left\langle G_R^{\lambda,\xi}(\mathbf{k}_{\parallel})\right\rangle\left\langle G_A^{\lambda,\xi}(\mathbf{k}_{\parallel})\right\rangle\mathbf{k}_{\parallel}\cdot\mathbf{k}_{\parallel}. \label{eq:conductivity_magnit_no_vertex}	
	    \end{align}

\subsection{Vertex corrections}
\begin{figure}[!ht]
		\centering
		\includegraphics[width=0.9 \columnwidth]{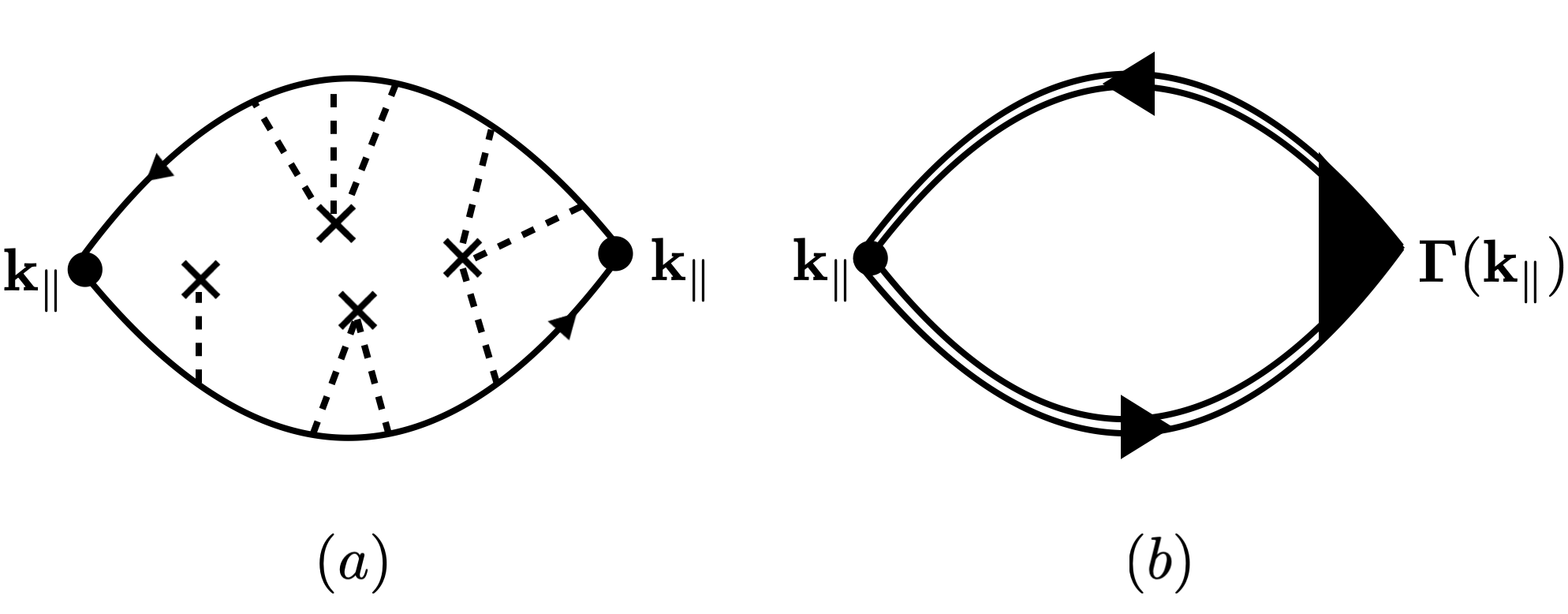}
		\caption{(a) A typical diagram contributing to the conductivity in Eq~\eqref{eq:conductivity_magnit_no_vertex}, involving the configurational average of the two internal GF with cross-links among them. The upper line corresponds to the retarded GF and the lower to the advanced GF. (b)  Diagrammatic representation of the two complete averaged GF (double lines) corresponding to the sum of all diagrams of the kind in (a) with the vertex correction $\boldsymbol{\Gamma}(\mathbf{k}_{\parallel})$.}
		\label{fig:vertex}
	\end{figure}

	The self-energy contribution modifies the definition of the retarded and advanced Green´s functions in Eq.~\eqref{eq:conductivity_magnit_no_vertex}, as depicted by the double lines in Fig.\ref{fig:vertex}(b). However, there are also scattering processes involving links between the two internal Green function lines, as depicted in FIG.~\ref{fig:vertex}(a). When considering such diagrams  with cross-links, as in FIG.~\ref{fig:vertex}(a), we must include the vertex correction as depicted in FIG.~\ref{fig:vertex}(b).
Taking into account the vertex correction, the conductivity becomes
	\begin{align}
	   \sigma_{\alpha\beta}^{(\lambda,\xi)}(T)&= \delta_{\alpha\beta}\frac{e^2\hbar v_{F}^2}{\pi^3 }\int_{0}^{\infty} dk \,\, \int_{-\infty}^{\infty}dE\,\, \left(-\frac{\partial f_0(E)}{\partial E}\right) \notag\\
    &\quad \times \left\langle G_R^{\lambda,\xi}(\mathbf{k}_{\parallel})\right\rangle\left\langle G_A^{\lambda,\xi}(\mathbf{k}_{\parallel})\right\rangle\mathbf{k}_{\parallel}\cdot\boldsymbol{\Gamma}_{RA}(\mathbf{k}_{\parallel},E), \label{eq:conductivity_magnit_vertex}	
	    \end{align}
	    \begin{figure}[!ht]
		\centering
		\includegraphics[width=1.0 \columnwidth]{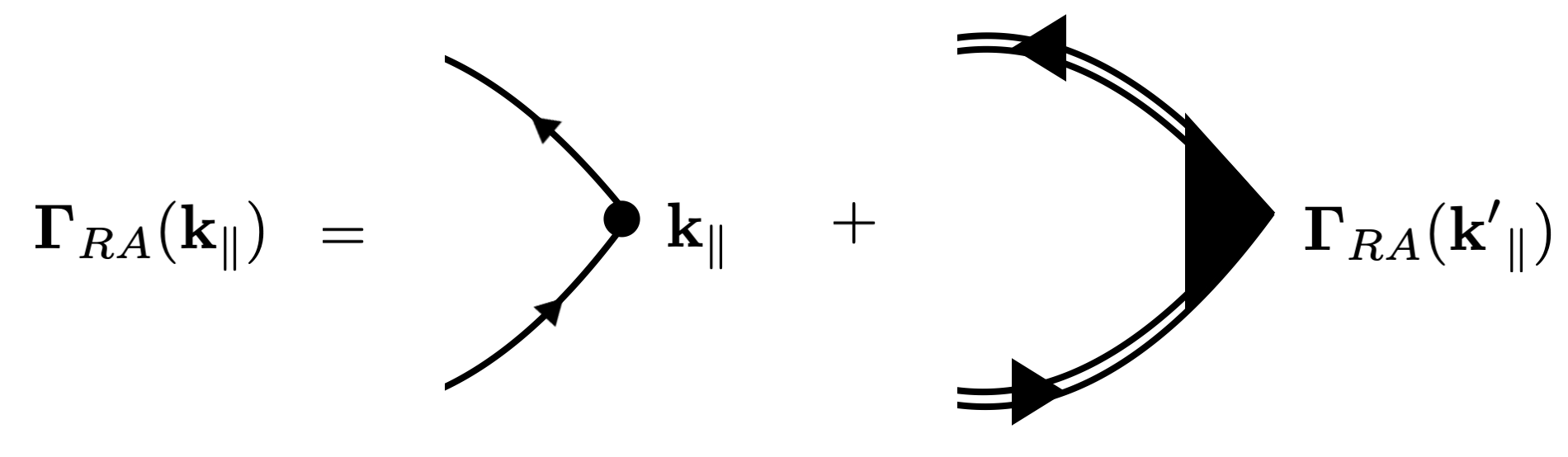}
		\caption{The Bethe-Salpeter integral equation for the vertex function $\boldsymbol{\Gamma}_{RA}(\mathbf{k}_{\parallel})$.}
		\label{fig:bethe_salpeter}
	\end{figure}
	where the vertex function $\boldsymbol{\Gamma}_{RA}(\mathbf{k}_{\parallel},E)$ is given as the solution to the \textit{Bethe-Salpeter equation} as depicted in FIG. \ref{fig:bethe_salpeter}. Then, we have
	  \begin{eqnarray}
		\boldsymbol{\Gamma}_{RA}(\mathbf{k}_{\parallel},E)&=&\mathbf{k}_{\parallel}+n_d\int \frac{d^2k'}{(2\pi)^2} \left\langle G_R^{\lambda,\xi}(\mathbf{k'}_{\parallel})\right\rangle\left\langle G_A^{\lambda,\xi}(\mathbf{k'}_{\parallel})\right\rangle\nonumber\\
		&&\times\left| T^{(\lambda,\xi)}_{\mathbf{k'}_{\parallel}\mathbf{k}_{\parallel}} \right|^2\boldsymbol{\Gamma}_{RA}(\mathbf{k'}_{\parallel},E). \label{eq:bethe_salpeter}
	    \end{eqnarray}
The iterative solution of the Eq.~\eqref{eq:bethe_salpeter} for $\boldsymbol{\Gamma}_{RA}(\mathbf{k}_{\parallel},E)$ shows that the vertex function must be of the form 
	\begin{equation}
	    \boldsymbol{\Gamma}_{RA}(\mathbf{k}_{\parallel},E)=\gamma(\mathbf{k}_{\parallel},E)\mathbf{k}_{\parallel}.\label{eq:gamma}
	\end{equation}
	Then we obtain an integral equation for the scalar function $\gamma(\mathbf{k}_{\parallel},E)$ that in the low concentration limit becomes
	\begin{eqnarray}
		\gamma(\mathbf{k}_{\parallel},E)&=&1+n_d\frac{2\pi}{\hbar}\int \frac{d^2k'}{(2\pi)^2}\tau^{(\lambda,\xi)}(k')\left| T^{(\lambda,\xi)}_{\mathbf{k'}_{\parallel}\mathbf{k}_{\parallel}} \right|^2\nonumber\\
		&&\times\delta(E-\lambda\xi\hbar v_{F} k')\,\,\gamma(\mathbf{k'}_{\parallel},E)\frac{\mathbf{k}_{\parallel}\cdot\mathbf{k'}_{\parallel}}{k^2}. \label{eq:eqn_gamma}
	    \end{eqnarray}
     In the limit of low concentrations, we use the result in Appendix \ref{app:linear}, Eq.\eqref{eq:limit_GRGA}, to obtain
	\begin{align}
	    \sigma_{\alpha\alpha}^{(\lambda,\xi)}(T)&=\frac{2e^2 v_{F}^2}{\pi^2 }\int_{0}^{\infty} dk\, k^2\, \left(-\frac{\partial f_0(E)}{\partial E}\right)_{E=\lambda\xi\hbar v_{F} k}\notag\\
         &\quad\times
     \tau^{(\lambda,\xi)}(k)\,\,\gamma(\mathbf{k}_{\parallel},\lambda\xi\hbar v_{F} k).\label{eq:conductivity_low_conc}
	\end{align}
At low temperatures, an exact solution is possible since the Fermi distribution derivative takes a compact support at the Fermi energy. Then we can evaluate $\gamma(k)$ and $\tau^{(\lambda,\xi)}(k)$ at the Fermi momentum $k_F^{\xi}$, to obtain
\begin{equation}
	    \gamma(k_F^{\xi})=\frac{\tau^{(\lambda,\xi)}_1(k_F^{\xi})}{\tau^{(\lambda,\xi)}_1(k_F^{\xi})-\tau^{(\lambda,\xi)}(k_F^{\xi})},
\end{equation}
where we defined (for $\cos\phi'=\mathbf{k}_{\parallel}\cdot\mathbf{k'}_{\parallel}/k^2$)
\begin{equation}
	    \frac{1}{\tau^{(\lambda,\xi)}_1(k_F^{\xi})}=n_d\frac{2\pi}{\hbar}\int \frac{d^2k'}{(2\pi)^2}\left| T^{(\lambda,\xi)}_{\mathbf{k'}_{\parallel}\mathbf{k}_{\parallel}} \right|^2\cos \phi'\,\,\delta(\hbar v_F k_F^{\xi}-\hbar v_F k'). \label{eq:tau_1}
\end{equation}
After the substitution in Eq.~\eqref{eq:conductivity_low_conc} of $\gamma(\mathbf{k}_{\parallel},E)$ given in Eq.~\eqref{eq:eqn_gamma}, we get
	\begin{eqnarray}
	\sigma_{\alpha\alpha}^{(\lambda,\xi)}(T)&=&\frac{2e^2 v_{F}^2}{\pi^2 k_B T}\tau^{(\lambda,\xi)}_{\text{tr}}(k_F^{\xi})\int_{0}^{\infty}dk \, k^2 \,  f_0\left(\mathcal{E}^{\lambda,\xi}_{\mathbf{k}_{\parallel}}\right) \left[1-f_0\left(\mathcal{E}^{\lambda,\xi}_{\mathbf{k}_{\parallel}}\right) \right]\nonumber\\
 	&=&-\frac{4}{\pi^2v_F}\left(\frac{e^2}{\hbar}\right)\left(\frac{k_B T}{\hbar}\right)^2\tau^{(\lambda,\xi)}_{\text{tr}}(k_F^{\xi})\,\text{Li}_2 \left(-e^{\frac{\hbar v_{F} k_F}{k_B T}}\right),\label{eq:conductivity_vertex_final}
	\end{eqnarray} 
 where $\text{Li}_2(x)$ is the polylogarithm of order 2. Here, the total \textit{transport relaxation time} is defined by
	\begin{eqnarray}
	    &&\frac{1}{\tau^{(\lambda,\xi)}_{\text{tr}}(k_F^{\xi})}=\frac{1}{\tau^{(\lambda,\xi)}(k_F^{\xi})}-\frac{1}{\tau^{(\lambda,\xi)}_1(k_F^{\xi})}\\
	    &=&\frac{2\pi n_d}{\hbar}\int \frac{d^2k'}{(2\pi)^2}\delta(\hbar v_F k_F^{\xi}-\hbar v_F k')\left| T^{(\lambda,\xi)}_{\mathbf{k'}_{\parallel}\mathbf{k}_{\parallel}} \right|^2(1-\cos \phi')\nonumber
	\end{eqnarray}
Using the form of the $T$-matrix elements in Eq.~\eqref{eq:Tmatrix_final}, we get a closed expression in terms of the scattering phase shifts $\delta_m(k)$
\begin{equation}
	    \frac{1}{\tau^{(\lambda,\xi)}_{\text{tr}}(k_F^{\xi})}=\frac{2 n_d v_F}{k_F^{\xi}}\sum_{m=-\infty}^{\infty}\sin^2 \left[\delta_{m}(k_F^{\xi})-\delta_{m-1}(k_F^{\xi}) \right].
	    \label{eq_tau_k}
\end{equation}
From Eq.~\eqref{eq:conductivity_vertex_final}, we can investigate the zero temperature $T\rightarrow 0$ and high temperature $T\gg \hbar v_F k_F/k_B$ limits, respectively. In the zero temperature limit, we obtain
\begin{equation}
		\sigma_{\alpha\alpha}^{(\lambda,\xi)}(T\rightarrow 0)=\frac{2}{\pi^2} k_F^{\xi 2}\left(\frac{e^2}{\hbar }\right)v_{F,\alpha}\tau^{(\lambda,\xi)}_{\text{tr}}(k_F^{\xi}),\label{eq:conductivity_Tcero_vertex}
\end{equation}
a constant that depends on the miscroscopic material properties (such as $v_F$), as well as on the concentration of dislocations $n_d$ through the relaxation time.

On the other hand, in the high-temperature limit $T\gg \hbar v_F k_F/k_B$, we obtain a quadratic dependence on temperature
\begin{equation}
		\sigma_{\alpha\alpha}^{(\lambda,\xi)}(T\gg \hbar v_F k_F/k_B)=\frac{1}{3v_{F,\alpha}}\left(\frac{e^2}{\hbar}\right)\left(\frac{k_B T}{\hbar}\right)^2\tau^{(\lambda,\xi)}_{\text{tr}}(k_F^{\xi}),\label{eq:conductivity_T_large_vertex}
\end{equation}
where the overall constant depends on the microscopic parameters for each material, as well as on the concentration of dislocations through the relaxation time.

\section{RESULTS}
	\label{sec:results}
	
	In this section, we apply the theory and analytical expressions obtained in the previous section to calculate the electrical conductivity of several materials in the family of transition metals monopnictides, i.e. TaAs, TaP, NbAs and NbP. For an estimation of the concentration of defects $n_d$ in real crystal systems, the Ref.~\cite{Szot_2018} reports that the native concentration of dislocations in the lattice of the materials TiO$_2$ and SrTiO$_3$, vary in the range $n_d \sim 10^{5} - 10^7\,\text{cm}^{-2}$. These concentrations can be enhanced using different treatments up to $10^{13}\,\text{cm}^{-2}$,  nearly to the rendering amorphous limit. The microscopic/atomistic parameters involved in our theory are obtained from \textit{ab-initio} studies for WSM materials, as reported in Ref.~\cite{Lee_2015} and Ref.~\cite{Grassano_2018}. In particular, the later reference identifies anisotropies in the Fermi velocities and density of charge carries at different Weyl nodes and bands. Using these results for the densities of carriers, we compute the Fermi momentum at each Weyl node, i.e., $k_F^{\xi}$, as displayed in TABLE~\ref{tab:kFs}.
	
	\begin{table}[!ht]
    \centering
		\begin{tabular}{||c | c |c||} 
			\hline
			Material & $k_F^{+}$ [nm$^{-1}$] & $k_F^{-}$ [nm$^{-1}$] \\ [0.5ex] 
			\hline\hline
			TaAs & 0.23 & 0.05 \\ 
			\hline
			TaP & 0.50  & 0.09 \\
			\hline
			NbAs & 0.46 & 0.03\\
			\hline
			NbP & 1.04 & 0.15 \\
			\hline
		\end{tabular}
		\caption{Values of $k_F^{\xi}$ computed from the carrier densities reported in Ref.~\cite{Grassano_2018}.}
		\label{tab:kFs}
	\end{table}
	
	In what follows, for definiteness we shall assume that the axis of the defects is along the crystallographic $z$-direction and that we are measuring the conductivity along the $x$-direction. Then, we use the reported $x$-components of the Fermi velocities\cite{Lee_2015,Grassano_2018}. We have different Fermi velocities $v^{(\lambda,\xi)}_{F,x}$, for the conduction band ($\lambda=+1$) and for the valence band ($\lambda=-1$), and for each of the Weyl nodes ($\xi=\pm$). Actually, for the valence band Refs.~\cite{Lee_2015,Grassano_2018} report the \textit{hole} velocity. Their results are presented in the TABLE \ref{tab:vFs}. 
	
	\begin{table}[!ht]
    \centering
		\begin{tabular}{||c | c |c|c|c||} 
			\hline
			Material & $v^{(+,+)}_{F,x}$  & $v^{(-,+)}_{F,x}$  & $v^{(+,-)}_{F,x}$  & $v^{(-,-)}_{F,x}$  \\ [0.5ex] 
			\hline\hline
			TaAs & 3.2 & -5.3 & 2.6 & -4.3 \\ 
			\hline
			TaP & 3.7  & -5.4 & 2.0 & -3.9 \\
			\hline
			NbAs & 3.0 & -4.8 & 2.5 & -3.2 \\
			\hline
			NbP & 3.0 & -5.1 & 1.7 & -2.4  \\
			\hline
		\end{tabular}
		\caption{Values of the Fermi velocity $v^{(\lambda,\xi)}_{F,x}$ in units of $10^5$ m/s, as reported in Ref.~\cite{Grassano_2018}. Notice that for the valence bands ($\lambda=-1$) they report the hole velocity. }
		\label{tab:vFs}
	\end{table}
	
	Now, in order to study the additional effect of the torsional dislocations, we follow our previous work~\cite{Bonilla_Munoz_2021}. We assume that the dislocations are cylindrical regions along the $z$-axis with radius $a$. Here, we further assume that the defects possess an average radius of $a=15$ nm.  The simple relation between the torsional angle $\theta$ (in degrees) and the pseudo-magnetic field representing strain is $B_S a^2=1.36\,\theta\, \tilde{\phi}_{0}$~\cite{Bonilla_Munoz_2021}, where the modified flux quantum in this material is approximately $\displaystyle\tilde{\phi}_{0}\equiv \frac{\hbar v_F}{e}=\frac{1}{2\pi}\frac{v_F}{c}\frac{hc}{e}=\frac{1}{2\pi}\frac{1.5}{300}\cdot4.14\times10^5$ T$\mathring{\text{A}}^2\approx330$ T$\mathring{\text{A}}^2$. In this work, we have chosen a torsion angle $\theta=15^{\circ}$. The lattice mismatch effect at the surface of the dislocation cylinders is modeled by a repulsive delta-potential, with strength $V_0$, expressed in terms of the ``spinor rotation'' angle $\alpha = V_0/\hbar v_F$. According to our previous work~\cite{Bonilla_Munoz_2021}, a realistic choice is $\alpha=3\pi/4$. 
	
	With all of these parameters fixed, we can compute the total relaxation time for each material.
	%\begin{equation}
	%    \tau=\sum_{\xi=\pm 1}\sum_{\lambda=\pm 1} \tau^{(\lambda,\xi)}(k_F^{\xi}),
	%\end{equation}
	%where $\tau^{(\lambda,\xi)}(k_F^{\xi})$ is given in Eq.\eqref{eq_tau_k}. In the same way, we can compute the total transport relaxation time given by the vertex correction method
	%\begin{equation}
	    %\tau_{\text{tr}}=\sum_{\xi=\pm 1}\sum_{\lambda=\pm 1} \tau_{\text{tr}}^{(\lambda,\xi)}(k_F^{\xi}),
	    %\label{eq_tau_tot}
	%\end{equation}
	%where $\tau_{\text{tr}}^{(\lambda,\xi)}(k_F^{\xi})$ is given in Eq.\eqref{eq_tau_k}. 
 Our results are presented in TABLE~\ref{tab:relaxation_times}.
	\begin{table}[!ht]
    \centering
		\begin{tabular}{||c | c |c||} 
			\hline
			Material & $\tau$ [$10^{-13}$ s] & $\tau_{\text{tr}}$ [$10^{-13}$ s]\\ [0.5ex] 
			\hline\hline
			TaAs & 2.2 & 2.6 \\ 
			\hline
			TaP & 2.4  & 3.2 \\
			\hline
			NbAs & 2.2 & 3.1 \\
			\hline
			NbP & 2.4 & 4.2 \\
			\hline
		\end{tabular}
		\caption{Computed values for the total relaxation time and the total transport relaxation time for each material.  We consider a concentration of dislocations $n_d=10^{11}$ cm$^{-2}$.}
		\label{tab:relaxation_times}
	\end{table}
	Now, we compute the conductivity along the $x$-direction $\sigma_{xx}$. In what follows, we simply call it $\sigma(T)$, as a function of temperature. The total conductivity is the sum over nodes and bands
	\begin{equation}
	    \sigma(T)=\sum_{\xi=\pm 1}\sum_{\lambda=\pm 1}\sigma_{xx}^{(\lambda,\xi)}(T),
	\end{equation}
	where $\sigma_{xx}^{(\lambda,\xi)}(T)$ is given in Eq.~\eqref{eq:conductivity_vertex_final}, including the vertex correction. Our results for $T=0$ are presented in the TABLE \ref{tab:conductivity_T0}.
	\begin{table}[!ht]
    \centering
		\begin{tabular}{||c |c||} 
			\hline
			Material & $\sigma_0$ [$10^{3}$ $\Omega^{-1}$ cm$^{-1}$] \\ [0.5ex] 
			\hline\hline
			TaAs &  1.5 \\ 
			\hline
			TaP &  7.9 \\
			\hline
			NbAs & 7.5 \\
			\hline
			NbP &  34.6 \\
			\hline
		\end{tabular}
		\caption{Computed values for the total conductivity $\sigma_0 = \sigma(T=0)$ at zero temperature for each material. We consider a value of $n_d=10^{11}$ cm$^{-2}$.}
		\label{tab:conductivity_T0}
	\end{table}
	
\begin{figure}[!ht]
		\centering
		\includegraphics[width=1 \columnwidth]{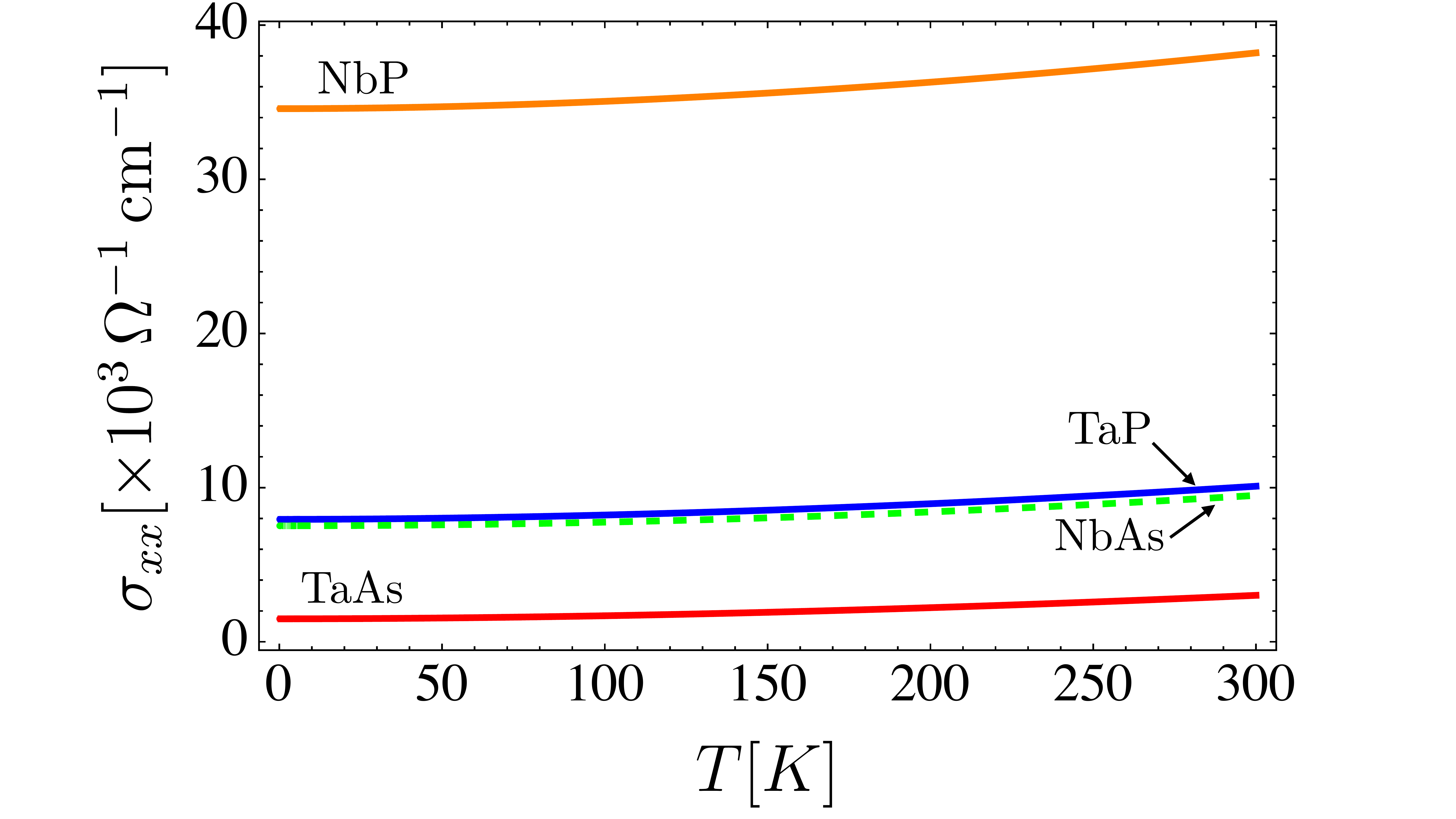}
		\caption{A comparison of the total conductivity versus temperature behavior for the transition metals monopnictides TaAs, TaP, NbAs and NbP. We use a value of $n_d=10^{11}$ cm$^{-2}$.  }
		\label{fig:G_T_all}
	\end{figure}

\begin{figure*}[!t]
	  \centering
      {\includegraphics[width=0.8\columnwidth]{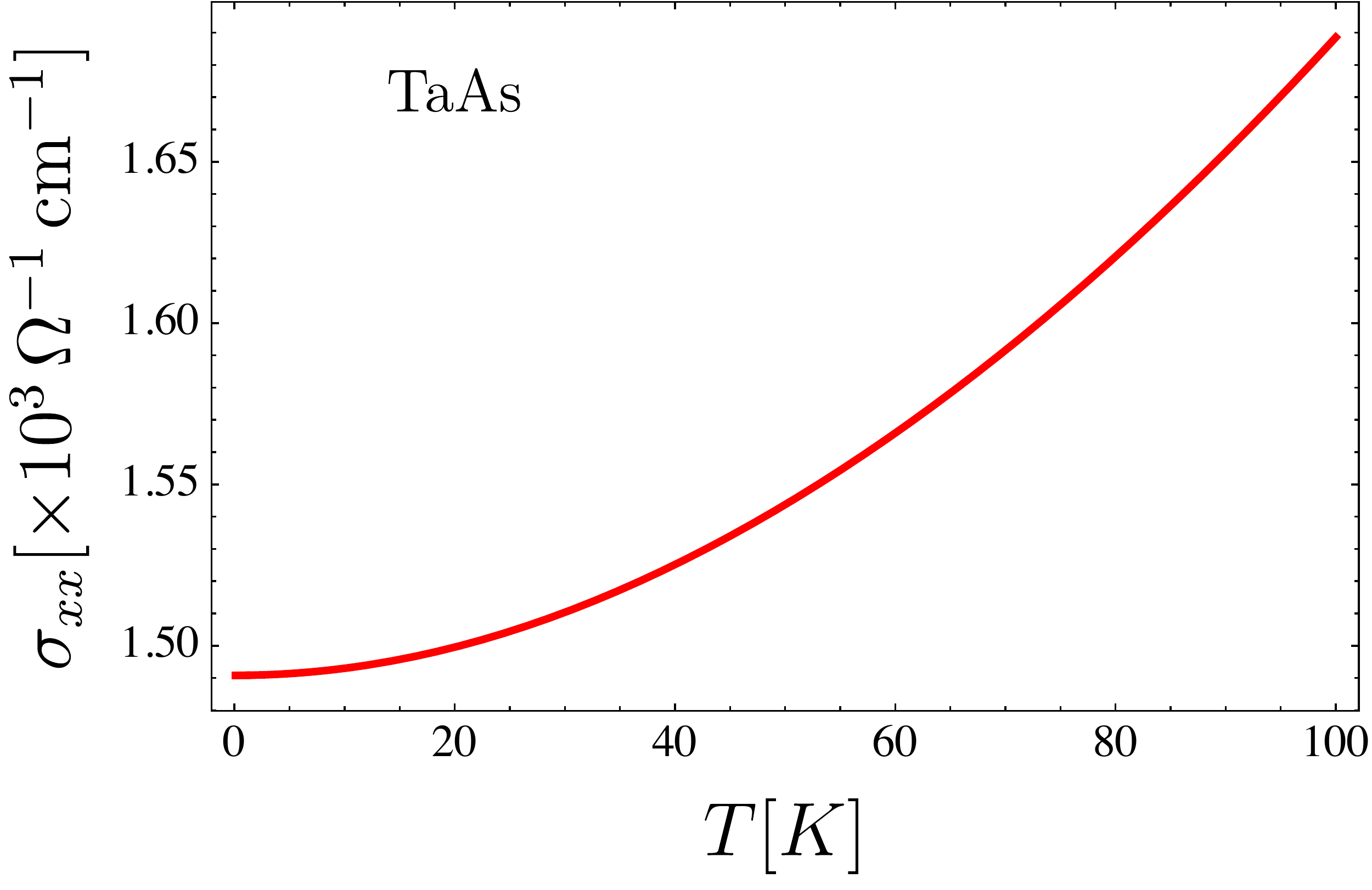}}
      \hskip .1cm
      {\includegraphics[width= 0.8\columnwidth]{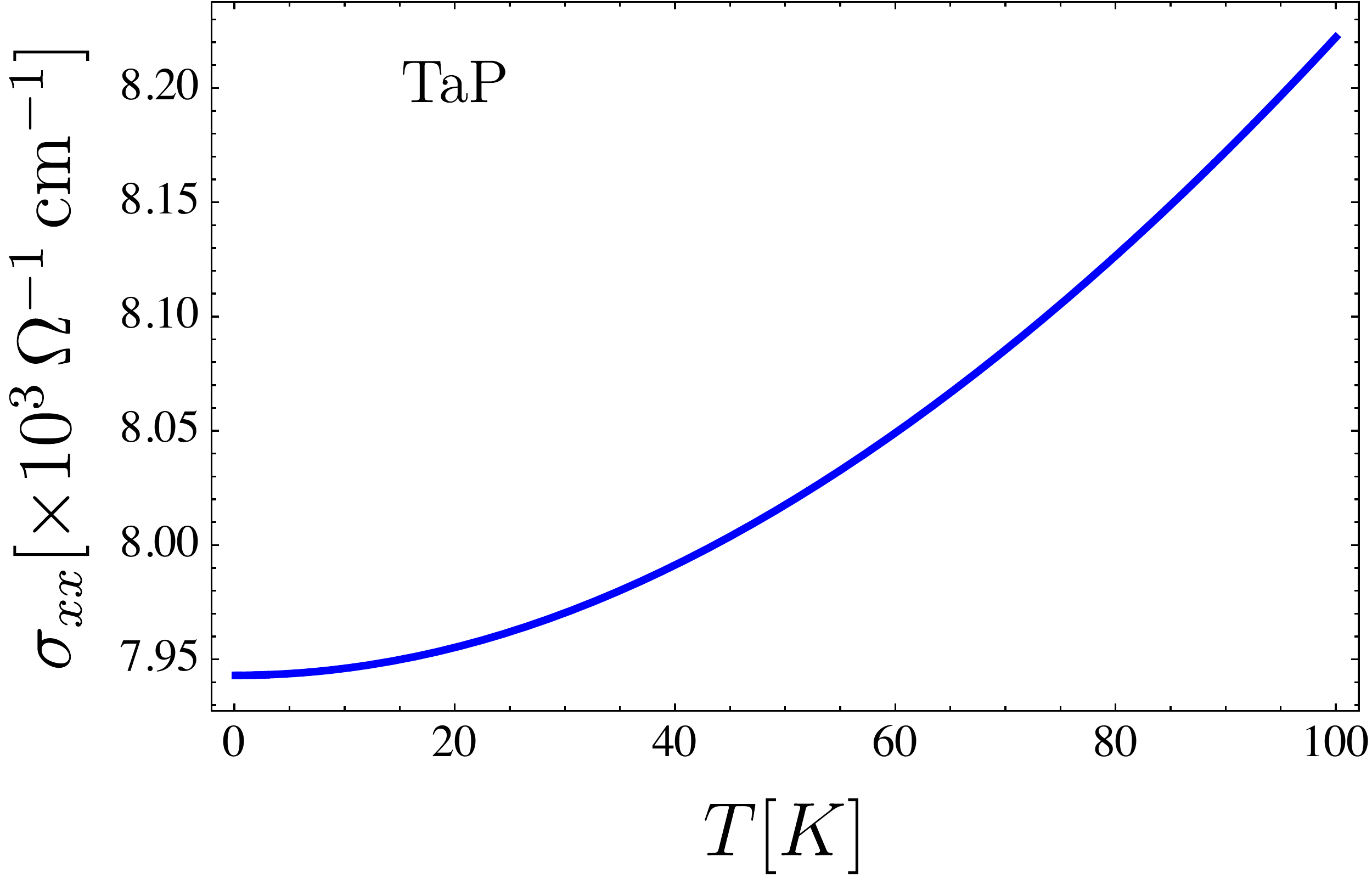}}
      \hskip .1cm
      {\includegraphics[width= 0.8\columnwidth]{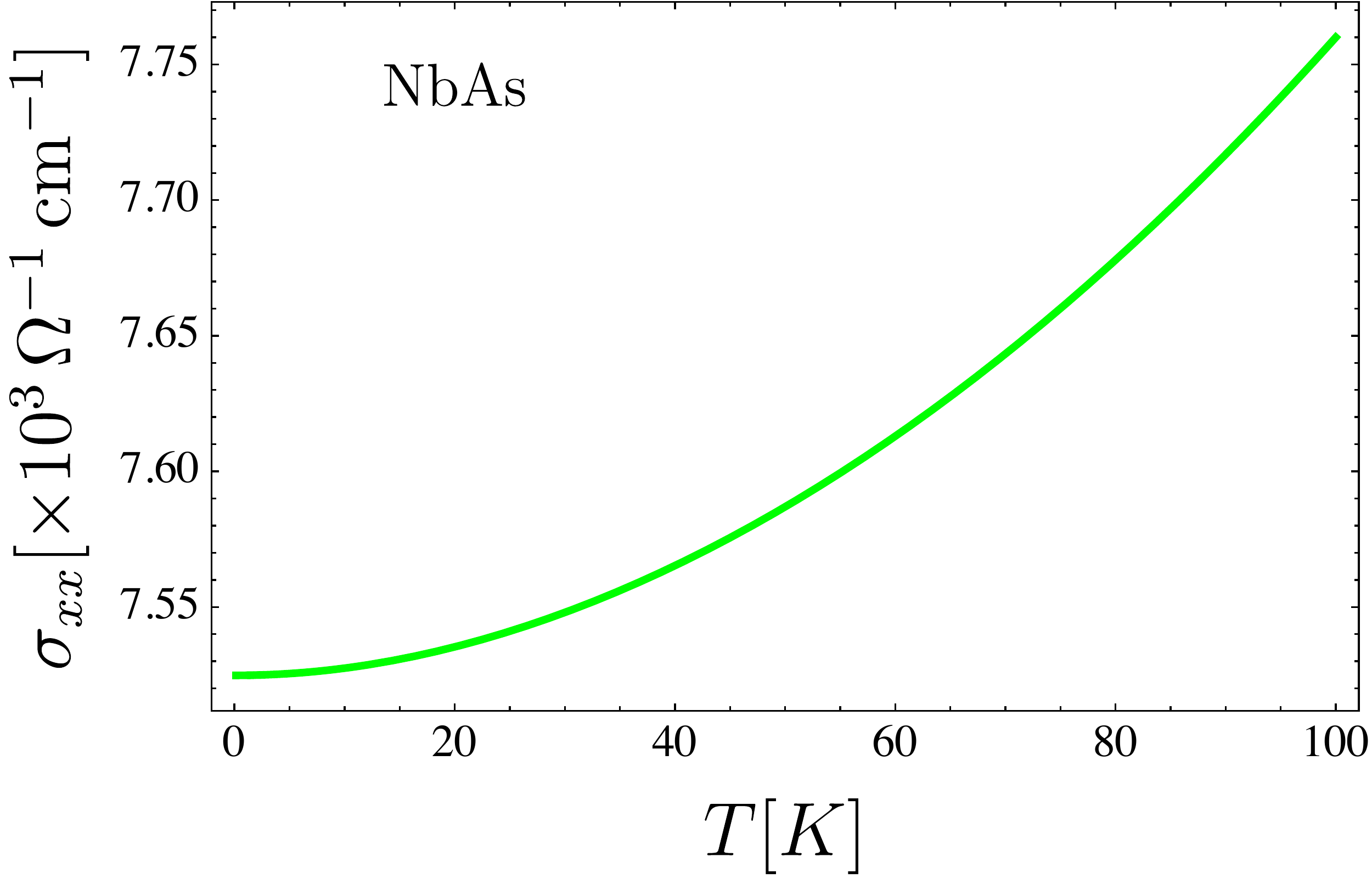}}
      \hskip .1cm
      {\includegraphics[width= 0.8\columnwidth]{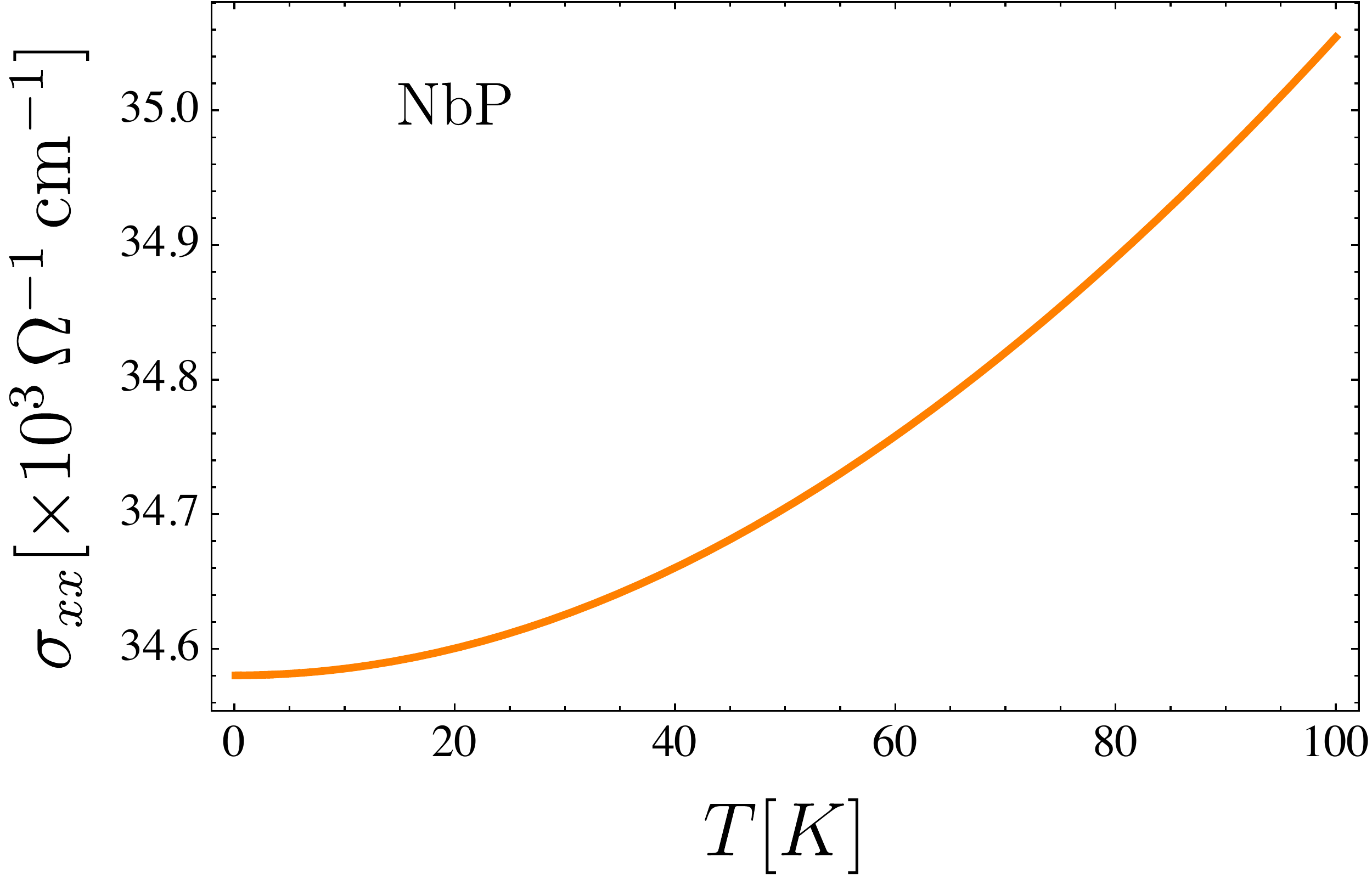}}
      \caption{The total conductivity vs. temperature behavior for the transition metals monopnictides TaAs, TaP, NbAs and NbP. We use a value of $n_d=10^{11}$ cm$^{-2}$..}
      \label{fig:G_vs_T}
	\end{figure*}

%	A plot comparing the behavior of the computed conductivity at finite temperature with and without the vertex correction is given in FIG.\ref{fig:vertex_vs_novertex}.
%	\begin{figure}[!ht]
%		\centering
%		\includegraphics[width=0.8 \columnwidth]{scattering_dislocations/vertex_vs_novertex.pdf}
%		\caption{Plot of the total conductivity versus temperature for the material TaAs. The red line corresponds to the conductivity computed from Eq\eqref{eq:conductivity_T_final} without the vertex correction. The blue line corresponds to the total conductivity computed from Eq.\eqref{eq:conductivity_vertex_final} with the vertex correction. We have used a value of $n_d=10^{11}$ cm$^{-2}$. }
%		\label{fig:vertex_vs_novertex}
%	\end{figure}
%	A plot showing the low temperature limit of the conductivity is presented in FIG.\ref{fig:G_vs_G0}. 
%	\begin{figure}[!ht]
%		\centering
%		\includegraphics[width=1 \columnwidth]{scattering_dislocations/G_vs_G0.pdf}
%		\caption{Plot of the relative value of the total conductivity $G/G_0$ respect to the $T=0$ limit versus temperature for the material TaAs. The red line represents the behavior of the finite temperature conductivity in Eq.\eqref{eq:conductivity_vertex_final} and the dashed blue line the $T=0$ limit. We use a value of $n_d=10^{11}$ cm$^{-2}$. }
%		\label{fig:G_vs_G0}
%	\end{figure}
	The conductivity as a function of temperature, for the transition metals monopnictides TaAs, TaP, NbAs and NbP, is presented in FIG.~\ref{fig:G_T_all} for all of them compared, and individually in the pannel FIG.~\ref{fig:G_vs_T} (a)-(d).

\begin{figure}[!ht]
		\centering
		\includegraphics[width=1 \columnwidth]{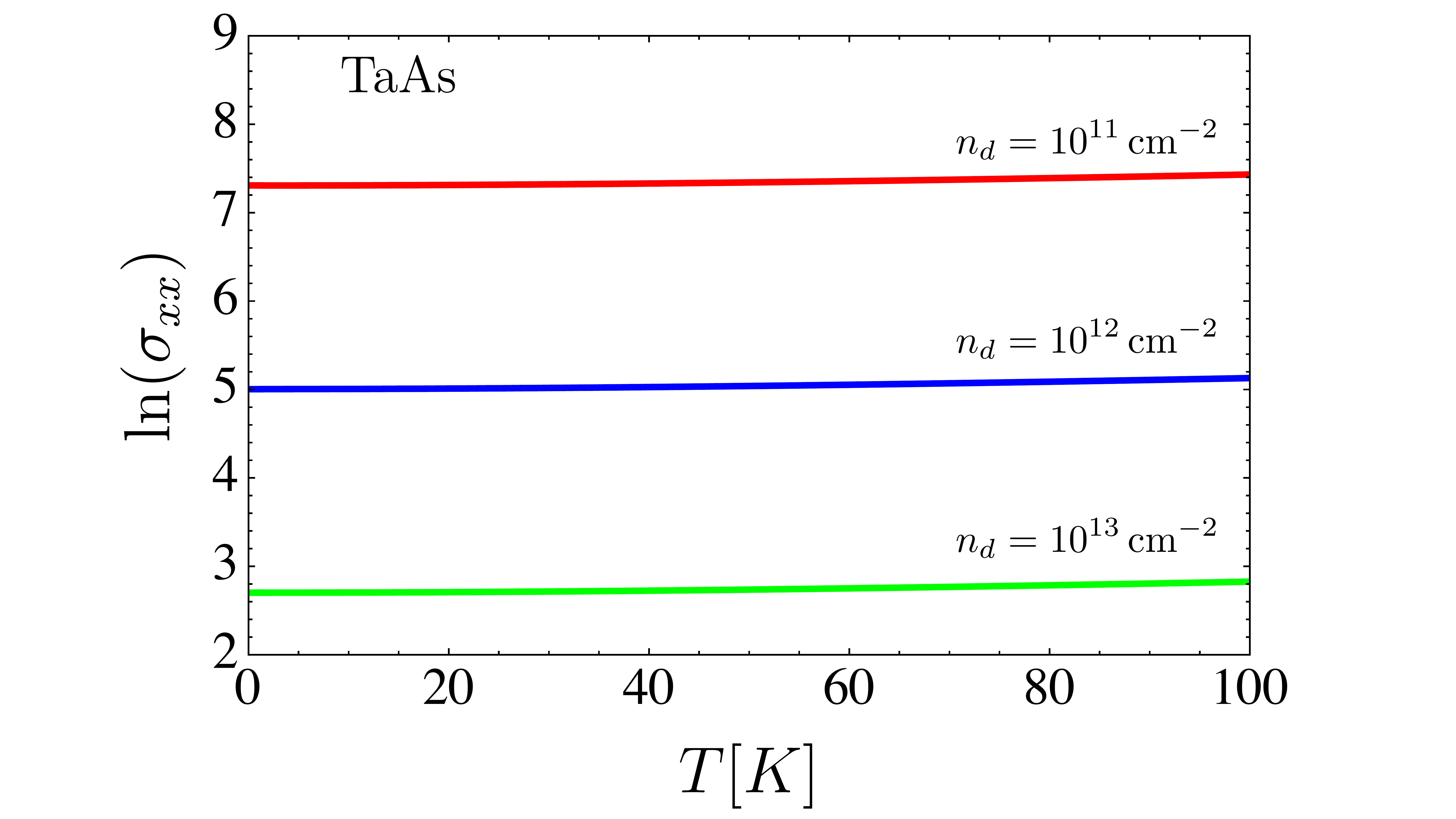}
		\caption{Natural logarithm of the conductivity versus temperature for 3 different concentrations of dislocations. The graphs were computed at zero temperature.}
		\label{fig:lnG_vs_T}
	\end{figure}

\begin{figure}[!ht]
		\centering
		\includegraphics[width=1 \columnwidth]{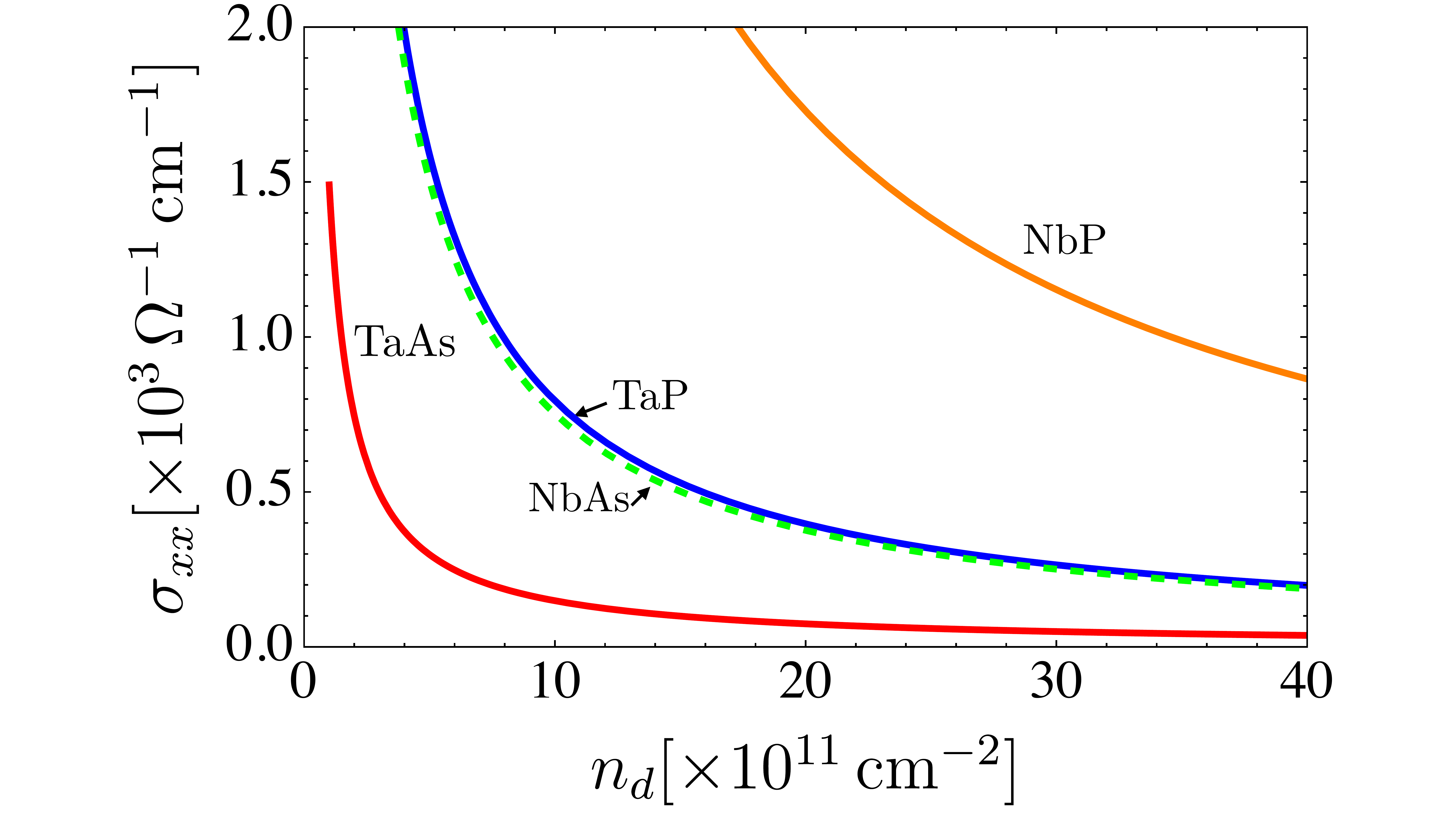}
		\caption{Plot of total conductivity versus defects' concentration. The graphs were computed at zero temperature.}
		\label{fig:G_vs_nd}
	\end{figure}

\begin{figure}[!ht]
		\centering
		\includegraphics[width=1 \columnwidth]{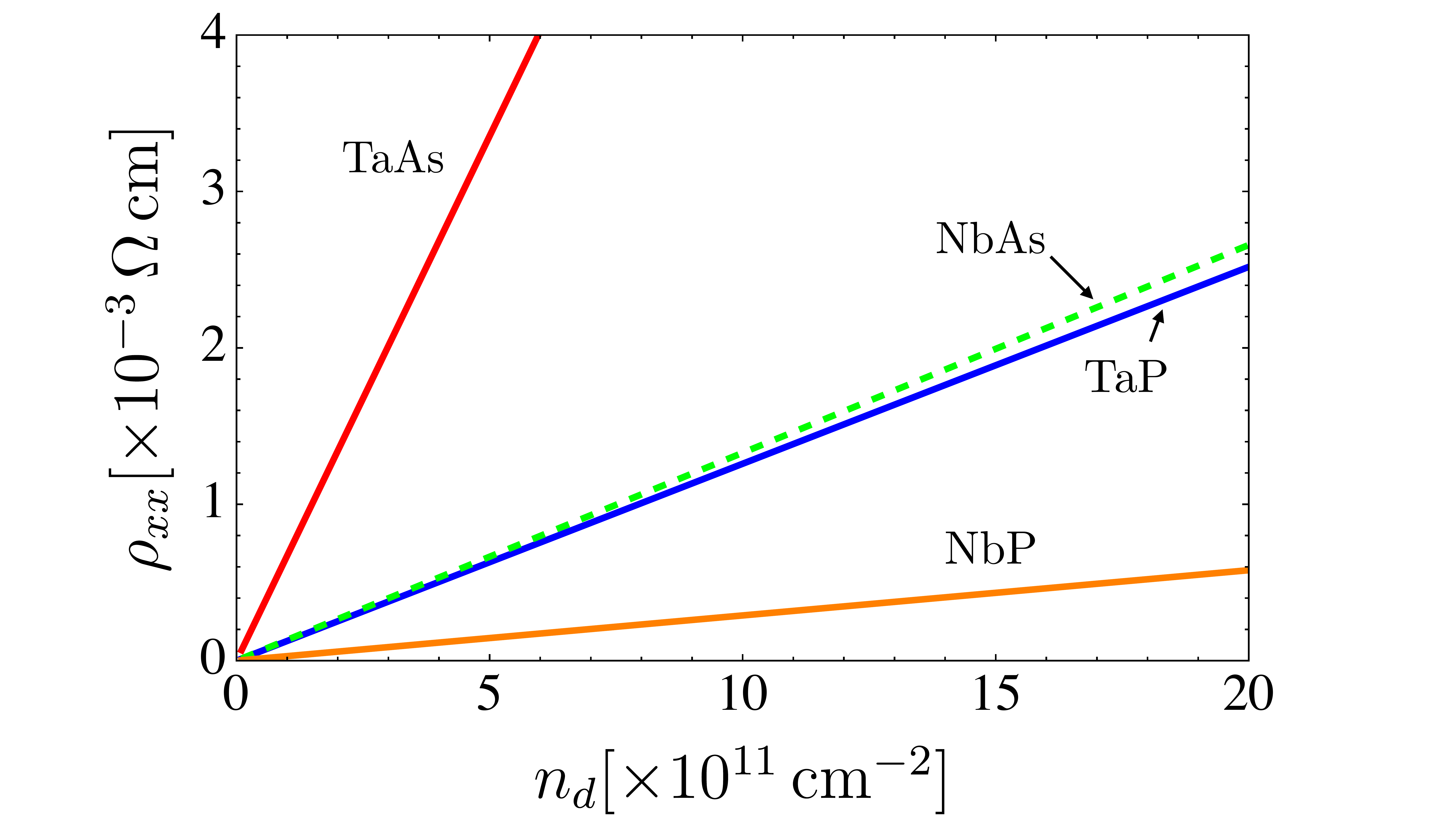}
		\caption{Total resistance, $R=1/G$, as a function of the concentration of defects $n_d$ for the family of materials TaAs, TaP, NbAs and NbP. The graphs were computed at zero temperature.}
		\label{fig:R_vs_nd}
	\end{figure}

%%%%%%%%%%%%%%%%%%%%%%%%%%%%%%%%%%%%%%%%%
	Now, let us study the conductivity behavior with respect to the density of dislocations $n_d$. In FIG.~\ref{fig:lnG_vs_T}, we present a plot of the natural logarithm of the conductivity versus temperature for three different concentrations of dislocations.
	The total conductivity as a function of the concentration of defects and at zero temperature is presented in FIG.~\ref{fig:G_vs_nd}. 
	Finally, a plot of the resistance, defined as the inverse of conductivity, as a function of the dislocations' density is presented in FIG.~\ref{fig:R_vs_nd}. 
\section{Conclusions}
In this work, we have studied the effect of a distribution of mechanical defects, i.e. torsional dislocations, over the electrical conductivity of the family of transition metals monopnictides TaAs, TaP, NbAs and NbP. Our theory is based on the mathematical analysis of the scattering phase shifts from a single defect, as stated in our previous work~\cite{Bonilla_Munoz_2021,Munoz2019,Munoz_2017,Soto_Garrido_2018,Soto_Garrido_2020}. We extended this previous analysis to develop a Green´s function formalism, in order to represent the scattering due to a finite concentration of randomly distributed defects. Within the non-crossing approximation for the self-energy, we solved explicitly for the disorder-averaged retarded Green´s function, that allows us to calculate the electrical conductivity in the Kubo linear-response formalism. We obtained general analytical expressions in terms of the parameters involved in the low-energy model representing the family of materials, and using the {\it{ab-initio}} estimations for such parameters, we provided a characterization of the conductivity as a function of temperature and concentration of defects for the transition metal monopnictides TaAs, TaP, NbAs and NbP. As a universal feature, we identified a $\sim T^2$ temperature dependence for $T \gg \hbar v_F k_F/k_B$, where the pre-factor depends on material-specific microscopic parameters as well as in the concentration of dislocations $n_d$ through the scattering relaxation time.
Our results do not involve the electron-phonon scattering effects, that will presumably contribute at higher temperatures, which is a subject of further study.  

%------------------------------------
\appendix
%------------------------------------

%------------------------------------
\section{ Calculation of the retarded free Green's function}
\label{app:GF}

The free retarded Green's function (GF) is represented in the coordinate basis as follows
	\begin{equation}
		\mathbf{G}^{\xi }_{R,0}(\mathbf{r},\mathbf{r'};E)= \left\langle \mathbf{r}\middle| \frac{1}{E-\hat{H}^{\xi}_0+ i\eta^{+}}\middle| \mathbf{r'}\right\rangle. 
	\end{equation}
In this basis, $\mathbf{G}^{\xi }_{R,0}(\mathbf{r},\mathbf{r'};E)$ satisfies the differential equation
	\begin{equation}
		\left(E+i\eta^{+}-\xi  v_F\boldsymbol{\sigma}\cdot \frac{\hbar}{i}\boldsymbol{\nabla }\right) 	\mathbf{G}^{\xi }_{R,0}(\mathbf{r},\mathbf{r'};E)=\sigma_0 \delta^{(3)}(\mathbf{r}-\mathbf{r'}). \label{eq:diff_eq_G_0}
	\end{equation}
where $\sigma_0$ is the $2\times2$ unit matrix. Let us introduce the scalar GF, $\mathcal{G}^{\xi}_{R,0}(\mathbf{r},\mathbf{r'})$,  by means of the expression
	\begin{equation}
		\mathbf{G}^{\xi }_{R,0}(\mathbf{r},\mathbf{r'};E)=\left(E+i\eta^{+}+\xi  v_F\boldsymbol{\sigma}\cdot \frac{\hbar}{i}\boldsymbol{\nabla} \right) \mathcal{G}^{\xi}_{R,0}(\mathbf{r},\mathbf{r'}).\label{eq:G0_scalar}
	\end{equation}
Bearing in mind that we are treating the elastic scattering problem, the energy of the out-state must be the same as those of the incident-free-particle state, i.e., $E=\lambda\xi\hbar v_F |\mathbf{k}|$. Then, the scalar GF satisfies the Helmholtz equation
	\begin{equation}
		\left(\nabla^2 +k^2+i\eta^{+}\right) \mathcal{G}^{\xi}_{R,0}(\mathbf{r},\mathbf{r'})=\frac{1}{\hbar^2 v_F^2} \delta^{(3)}(\mathbf{r}-\mathbf{r'}). \label{eq:Helmholtz_eqn}
	\end{equation}
Due to the symmetry along the $z$-axis we can decouple it from its perpendicular plane as follows
		\begin{equation}
			\mathcal{G}^{\xi}_{R,0}(\mathbf{r},\mathbf{r'};k)=\int_{-\infty}^{\infty}\frac{dq_z}{2\pi}e^{iq_z (z-z')}\mathcal{G}^{\xi}_{R,0}(\mathbf{x},\mathbf{x'};q_z,k),\label{eq:G_q_z}
		\end{equation}
where $\mathcal{G}^{\xi}_{R,0}(\mathbf{x},\mathbf{x'};q_z,k)$ is a reduced GF and $\mathbf{x}=(x,y)$ is the position vector on the plane. Then, the Helmholtz equation for the reduced GF on the plane takes the form
   \begin{equation}
   \left(\nabla_{\parallel}^2 -q_z^2+k^2+i\eta^{+}\right) \mathcal{G}^{\xi}_{R,0}(\mathbf{x},\mathbf{x'};q_z,k)=\frac{1}{\hbar^2 v_F^2} \delta^{(2)}(\mathbf{x}-\mathbf{x'}), \label{eq:Helmholtz_2}
   \end{equation}
   where $\nabla_{\parallel}^2 =\partial_{x}^2+\partial_{y}^2$. As we can be seen from the FIG.\ref{fig:single_defect}, the free incident particle's propagation is normal to the cylinder's axis. We assume that the incident particles have negligible momentum along the $z$-axis, and by momentum conservation, they remain with negligible momentum along that direction during the transport process. Then, we can write $\mathbf{k}=(\mathbf{k}_{\parallel},0)$ where $\mathbf{k}_{\parallel}=(k_x,k_y)$. Hence, the system is reduced to an effective two-dimensional description and we can consider the reduced GF on the plane as independent of the Fourier mode $q_z$. Then, from Eq.\eqref{eq:G_q_z} we have $\mathcal{G}^{\xi}_{R,0}(\mathbf{r},\mathbf{r'};k)=\delta(z-z')\mathcal{G}^{\xi}_{R,0}(\mathbf{x},\mathbf{x'};k)$, and we can expand the reduced GF on the plane in the traverse Fourier space
  \begin{equation}
  	\mathcal{G}^{\xi}_{R,0}(\mathbf{x},\mathbf{x'};k)=\int \frac{d^2q_{\parallel}}{(2\pi)^2}e^{i\mathbf{q}_{\parallel}\cdot(\mathbf{x}-\mathbf{x'})}\tilde{\mathcal{G}}^{\xi}_{R,0}(\mathbf{q}_{\parallel}),\label{eq:G_Fourier}
  \end{equation}
  where $\mathbf{q}_{\parallel}=(q_x,q_y)$.  Replacing in the Eq.\eqref{eq:Helmholtz_2}, we obtain in the traverse Fourier space
		\begin{equation}
			\tilde{\mathcal{G}}^{\xi}_{R,0}(\mathbf{q}_{\parallel})=-\frac{1}{\hbar^2 v_F^2}\frac{1}{q_{\parallel}^2-k^2-i\eta^{+}}.
		\end{equation}
We perform the integration in Eq.\eqref{eq:G_Fourier} in polar coordinates
		\begin{align}
			\mathcal{G}^{\xi}_{R,0}(\mathbf{x},\mathbf{x'};k)
		&=-\frac{1}{\hbar^2 v_F^2}\frac{2\pi}{(2\pi)^2}\int_{0}^{\infty}dq_{\parallel}\, q_{\parallel} \frac{ J_0(q_{\parallel}R)}{q_{\parallel}^2-k^2-i\eta^{+}},
	\end{align}
	where $R=|\mathbf{x}-\mathbf{x'}|$ and we have used the integral representation of the Bessel functions $\int_{0}^{2\pi}e^{iz\cos\phi\pm i n\phi}d\phi=2\pi i^{n}J_n(z)$.
	In order to perform the last integration we need the result
	\begin{equation}
		\int_{0}^{\infty}\frac{x^{\nu+1}J_{\nu}(bx)}{\left(x^2+a^2 \right)^{\mu+1} }dx=\frac{a^{\nu-\mu}b^{\mu}}{2^{\mu}\Gamma(\mu+1)}K_{\nu-\mu}(ab),
	\end{equation}
	together with the relation $K_{n}(z)=\frac{\pi}{2}i^{n+1} H^{(1)}_n \left(iz \right)$. The result is
	\begin{equation}
		\mathcal{G}^{\xi}_{R,0}(\mathbf{x},\mathbf{x'};k)=-\frac{i}{4\hbar^2 v_F^2}H^{(1)}_0 \left(k|\mathbf{x}-\mathbf{x'}| \right) .
	\end{equation}
	This form is adequate because in the asymptotic form for large $|\mathbf{x}-\mathbf{x'}|$ it produces outgoing cylindrical waves as it is desired for the retarded GF. Now, to obtain the final form for the free GF matrix we apply the definition in Eq.\eqref{eq:G0_scalar} with $E=\lambda\xi\hbar v_F |\mathbf{k}|$, taking into account that we have reduced to a two dimensional system on the plane $x$-$y$
	\begin{align}
		\mathbf{G}^{\xi }_{R,0}\left( \mathbf{r},\mathbf{r'};k\right) &=-\frac{i \lambda \xi}{4\hbar v_F}\delta(z-z') \nonumber \\
 &\quad \quad\times \left( k \sigma_0-i\lambda\boldsymbol{\sigma}\cdot \boldsymbol{\nabla }_{\parallel}\right) H^{(1)}_0 \left(k|\mathbf{x}-\mathbf{x'}| \right). 
	\end{align}
In plane polar coordinates
	\begin{equation}
		\boldsymbol{\sigma}\cdot\boldsymbol{\nabla }_{\parallel}=\left( \boldsymbol{\sigma}\cdot \mathbf{\hat{r}} \right)\frac{\partial}{\partial r}+\left(\boldsymbol{\sigma}\cdot \boldsymbol{\hat{\varphi}}\right)\frac{1}{r}\frac{\partial}{\partial \varphi}
	\end{equation}
	where $r=|\mathbf{x}-\mathbf{x'}|$, $\varphi$ is the angle the vector $\mathbf{x}-\mathbf{x'}$ makes with the $x$ axis and
	\begin{equation}
		\boldsymbol{\sigma}\cdot \mathbf{\hat{r}}=\begin{pmatrix}
			0 & e^{-i\varphi} \\
			e^{i\varphi} & 0
		\end{pmatrix} , \quad \boldsymbol{\sigma}\cdot \boldsymbol{\hat{\varphi}}=\begin{pmatrix}
			0 & -i e^{-i\varphi} \\
			ie^{i\varphi} & 0
		\end{pmatrix} .
	\end{equation}
	The final form for the retarded Green's function matrix in the coordinates representation is 
	\begin{align}
		\mathbf{G}^{\xi }_{R,0}\left( \mathbf{r},\mathbf{r'};k\right)&= -\frac{\lambda \xi ik}{4 \hbar v_F}\delta(z-z')\nonumber \\
  &\times \begin{bmatrix}
			H^{(1)}_0 \left(k|\mathbf{x}-\mathbf{x'}| \right)& i\lambda e^{-i\varphi} H^{(1)}_1 \left(k|\mathbf{x}-\mathbf{x'}|\right) \\
			i\lambda e^{i\varphi} H^{(1)}_1 \left(k|\mathbf{x}-\mathbf{x'}|\right)& H^{(1)}_0 \left(k|\mathbf{x}-\mathbf{x'}| \right)
		\end{bmatrix}, \label{eq:GF_R}
	\end{align}
which produces Eq.\eqref{eq:G0_final}.

%------------------------------------
\section{Scattering by a single cylindrical defect}
\label{app:scattering}

We can represent the Lippmann-Schwinger Eq.\eqref{eq:Lippmann_Schwinger} in the coordinate basis as follows
		\begin{eqnarray}
		\Psi_{\mathbf{k},\lambda}(\mathbf{r}) = \Phi_{\mathbf{k},\lambda}(\mathbf{r})   +&&\int d^3r' \int d^3r'' \left\langle\mathbf{r}\middle|\hat{G}^{\xi}_{R,0}\left( E \right)\middle|\mathbf{r'}\right\rangle\nonumber\\
				&&\times\left\langle\mathbf{r'}\middle|\hat{T}^{\xi}(E)\middle|\mathbf{r''}\right\rangle \Phi_{\mathbf{k},\lambda}(\mathbf{r''}). \label{eq:Lippmann}
	\end{eqnarray}
The form of free spinors in Eq.\eqref{eq:spinor_final} with momentum $\mathbf{k}_{\parallel}$ on the $x$-$y$ plane is 
     	\begin{equation}
     		\Phi_{\lambda,\mathbf{k}}(\mathbf{r})=\frac{1}{\sqrt{2 }} \begin{pmatrix}
     			1 \\ \lambda e^{i\phi}
     		\end{pmatrix}e^{i\mathbf{k}\cdot \mathbf{r}}\equiv \frac{1}{\sqrt{2 }} \begin{pmatrix}
     			1 \\ \lambda e^{i\phi}
     		\end{pmatrix}e^{i\mathbf{k}_{\parallel}\cdot \mathbf{x}}=\Phi_{\lambda,\mathbf{k}_{\parallel}}(\mathbf{x}).\label{eq:free_spinor_plane}
     	\end{equation}
     where $k_x=k\cos \phi$ and $k_y=k\sin\phi$. The incident spinors are assumed to enter the scattering region with momentum along the $x$ axis, and they are represented by
    	\begin{equation}
    		\Phi^{\text{inc}}_{\lambda,\mathbf{k}_{\parallel}}(\mathbf{x})=\frac{1}{\sqrt{2 }} \begin{pmatrix}
    			1 \\ \lambda 
    		\end{pmatrix}e^{ikx}.\label{eq:inc_spinor}
    	\end{equation}	
	Now, $\hat{H}_1^{\xi}$ is a local potential independent of the $z$ coordinate as can be seen from Eq.\eqref{eq:H_1}. Then, the $T$-matrix is diagonal in the coordinate basis and depends only on vectors $\mathbf{x}$ on the plane. Thus, $\left\langle\mathbf{r'}\middle|\hat{T}^{\xi}(E)\middle|\mathbf{r''}\right\rangle=\mathbf{T}^{\xi}(\mathbf{x'},E)\delta^{(3)}(\mathbf{r'}-\mathbf{r''})$, where $\mathbf{T}^{\xi}(\mathbf{x'},E)$ is a $2\times2$ matrix. The incident spinor is given in Eq.\eqref{eq:inc_spinor}, and using the retarded GF in Eq.\eqref{eq:GF_R}, the Lippmann-Schwinger equation in Eq.\eqref{eq:Lippmann} is reduced to the $x$-$y$ plane as follows
	\begin{align}
		\Psi_{\mathbf{k}_{\parallel},\lambda}(\mathbf{x})&= \frac{1}{\sqrt{2}} \begin{pmatrix}
			1 \\ \lambda 
		\end{pmatrix}e^{ikx}\notag\\
  &\quad +\int d^2x' \mathbf{G}^{\xi }_{R,0}\left( \mathbf{x},\mathbf{x'};k\right)\mathbf{T}^{\xi}(\mathbf{x'},E)\Phi_{\mathbf{k}_{\parallel},\lambda}(\mathbf{x'}), \label{eq:Integral_Eq_T}
	\end{align}
 where $\mathbf{G}^{\xi }_{R,0}\left( \mathbf{x},\mathbf{x'};k\right)$ is given in Eq.\eqref{eq:G0_final}. The asymptotic form for large argument of the Hankel's functions are
	\begin{align}
		H^{(1)}_0(k|\mathbf{x}-\mathbf{x'}|)&\sim \sqrt{\frac{2}{i\pi k |\mathbf{x}-\mathbf{x'}| }}e^{ik|\mathbf{x}-\mathbf{x'}|}, \\
		H^{(1)}_1(k|\mathbf{x}-\mathbf{x'}|)&\sim -i\sqrt{\frac{2}{i\pi k |\mathbf{x}-\mathbf{x'}| }}e^{ik|\mathbf{x}-\mathbf{x'}|},
	\end{align}
	where we have used the known limiting form
	\begin{equation}
		H^{(1)}_{\nu}(x)\sim \sqrt{\frac{2}{\pi x}}e^{i\left(x-\frac{\nu \pi}{2}-\frac{\pi}{4} \right) }, \quad x\rightarrow \infty.
	\end{equation}
	Now, recall the geometry of the scattering process as depicted in FIG.\ref{fig:scattering}. We expand $|\mathbf{x}-\mathbf{x'}|$ for large $|\mathbf{x}|$ as follows
	\begin{equation}
		|\mathbf{x}-\mathbf{x'}|\sim r -\mathbf{\hat{n}}\cdot \mathbf{x'}+\mathcal{O}\left( (r'/r)^2\right),
	\end{equation}
	where $r=|\mathbf{x}|$, $r'=|\mathbf{x'}|$ and $\mathbf{\hat{n}}$ is the unit vector in the direction of $\mathbf{x}$, i.e., $\mathbf{\hat{n}}=\mathbf{x}/r$. Noting that in this asymptotic form the direction of $\mathbf{k'}_{\parallel}$ coincides with that of $\mathbf{x}$ and is practically  the same of $\mathbf{x}-\mathbf{x'}$, i.e., $\mathbf{k'}_{\parallel}=k\mathbf{\hat{n}}$ and that the angle $\phi$ the vector $\mathbf{k'}_{\parallel}$ makes with the $\mathbf{k}_{\parallel}$ incident momentum is approximately the same angle $\mathbf{x}-\mathbf{x'}$ does, i.e., $\phi\sim \varphi$,  we have the asymptotic form for the free Green's function in Eq.\eqref{eq:G0_final} 
	\begin{equation}
		\mathbf{G}^{\xi }_{R,0}\left( \mathbf{x},\mathbf{x'};k\right)\sim -\frac{\lambda \xi  k}{4\hbar v_F}\sqrt{\frac{2i}{\pi k}}\begin{bmatrix}
			1& \lambda e^{-i\phi}  \\
			\lambda e^{i\phi} & 1
		\end{bmatrix}e^{-i\mathbf{k'}_{\parallel}\cdot\mathbf{x'}}\frac{e^{ikr}}{\sqrt{r}}. \label{eq:G0_asymptotic}
	\end{equation}
	Replacing the asymptotic form in Eq.\eqref{eq:G0_asymptotic} in the Eq.\eqref{eq:Integral_Eq_T} we obatain Eq.\eqref{eq:assymptotic_wave_fn}
	\begin{equation}
		\Psi_{\mathbf{k}_{\parallel},\lambda}(\mathbf{x})\sim\frac{1}{\sqrt{2}} \begin{pmatrix}
			1 \\ \lambda 
		\end{pmatrix}e^{ikx}-\frac{\lambda \xi }{2 \hbar v_F}\sqrt{\frac{i k}{\pi}}T^{(\lambda,\xi)}_{\mathbf{k'}_{\parallel}\mathbf{k}_{\parallel}} \begin{pmatrix}
			1 \\ \lambda e^{i\phi}
		\end{pmatrix}\frac{e^{ikr}}{\sqrt{r}},
	\end{equation}
  where the $T$-matrix elements are
	\begin{equation}
		T^{(\lambda,\xi)}_{\mathbf{k'}_{\parallel}\mathbf{k}_{\parallel}}\left(E \right) =\int d^2x' \Phi^{\dagger}_{\mathbf{k'}_{\parallel},\lambda}(\mathbf{x'})\mathbf{T}^{\xi}(\mathbf{x'},E)\Phi_{\mathbf{k}_{\parallel},\lambda}(\mathbf{x'}) .\label{eq:T_mat_elements}
	\end{equation}
In order to compute the $T$-matrix elements we need the phase shifts, whose analytical expression is presented in Eq.(32) of the supplemental material of our previous work Ref.\cite{Bonilla_Munoz_2021}. Here we reproduce the final result
\begin{widetext}
\begin{equation}
		\tan \delta_{m}(k)= \frac{\beta J_{m+1}(ka) - \varrho_{n}^{\xi} J_{m}(ka)\cdot z_a^{\frac{|m+1|-|m|}{2}}\frac{L_{n'_{\rho}}^{|m+1|}\left( z_a\right)}{L_{n_{\rho}}^{|m|}\left( z_a\right)}+\tan \alpha \left[ J_{m}(ka)+\beta\varrho_{n}^{\xi}J_{m+1}(ka)\cdot z_a^{\frac{|m+1|-|m|}{2}}\frac{L_{n'_{\rho}}^{|m+1|}\left( z_a\right)}{L_{n_{\rho}}^{|m|}\left( z_a \right)}\right] }{\beta Y_{m+1}(ka) - \varrho_{n}^{\xi} Y_{m}(ka)\cdot z_a^{\frac{|m+1|-|m|}{2}}\frac{L_{n'_{\rho}}^{|m+1|}\left( z_a\right)}{L_{n_{\rho}}^{|m|}\left( z_a\right)}+\tan \alpha \left[ Y_{m}(ka)+\beta\varrho_{n}^{\xi}Y_{m+1}(ka)\cdot z_a^{\frac{|m+1|-|m|}{2}}\frac{L_{n'_{\rho}}^{|m+1|}\left( z_a\right)}{L_{n_{\rho}}^{|m|}\left( z_a \right)}\right] },\label{eq:phase_shift}
	\end{equation}
 \end{widetext}
where $z_a=|B_{\xi}|a^2/2\tilde{\phi}_0$ ($a$ is the cylinder's radius).

%------------------------------------
\section{The spectral function}
\label{app:spectral}

The spectral function can be defined as follows
\begin{eqnarray}
	\hat{\mathcal{A}}^{\xi}(E)&=&i\left[\hat{G}^{\xi}_{R}(E)-\hat{G}^{\xi}_{A}(E) \right],\label{eq:A}
\end{eqnarray}
in terms of the complete retarded and advanced Green's functions. Then, the spectral function is Hermitian $\left[ \hat{\mathcal{A}}^{\xi}(E)\right]^{\dagger}=\hat{\mathcal{A}}^{\xi}(E)$. Given the averaged complete retarded Green's function in Eq.\eqref{eq:GR_final}, the form of the spectral function in momentum space is
\begin{align}
	&\mathcal{A}^{\lambda,\xi}(\mathbf{k}_{\parallel})= i\left[ 	\left\langle G_R^{\lambda,\xi}(\mathbf{k}_{\parallel})\right\rangle-	\left\langle G_A^{\lambda,\xi}(\mathbf{k}_{\parallel})\right\rangle\right] \notag \\
	&= i\left[ \frac{1}{ E-\lambda\xi\hbar v_F k-\Sigma_R^{\lambda,\xi}(\mathbf{k}_{\parallel})}-\frac{1}{ E-\lambda\xi\hbar v_F k-\Sigma_A^{\lambda,\xi}(\mathbf{k}_{\parallel})}\right].
\end{align}
Clearly, it takes the form of a Lorentzian distribution with compact support around the free particle's energy
	\begin{equation}
		\mathcal{A}^{\lambda,\xi}(k)=  \frac{2\left(\frac{\hbar}{2\tau^{(\lambda,\xi)}(k)}  \right) }{\left(E-\mathcal{E}^{\lambda,\xi}_{k}\right) ^2+\left(\frac{\hbar}{2\tau^{(\lambda,\xi)}(k)} \right) ^2},
	\end{equation}
	where $\tau^{(\lambda,\xi)}(k)$ is the \textit{relaxation time} and $\mathcal{E}^{\lambda,\xi}_{k}=\lambda\xi\hbar v_F k$. In the limit of low  concentration of defects, i.e., large relaxation time because of Eq.\eqref{eq:tau}, the spectral function becomes a delta distribution
 \begin{equation}
     \lim_{\tau\rightarrow\infty}\mathcal{A}^{\lambda,\xi}(k)=2\pi\delta\left(E-\mathcal{E}^{\lambda,\xi}_{k}\right).\label{eq:A_delta}
 \end{equation}
 Due to its behavior as a Lorentzian, the spectral function has the important property \cite{Mahan}
	\begin{equation}
	     \mathcal{A}^{\lambda,\xi}(k) \mathcal{A}^{\lambda',\xi}(k')= \mathcal{A}^{\lambda,\xi}(k) \mathcal{A}^{\lambda,\xi}(k)\delta_{\lambda\lambda'}.\label{eq:A^2}
	\end{equation}
Representing the spectral function Eq.\eqref{eq:A} in the coordinate basis using the complete set of eigenstates of the full Hamiltonian we have
	\begin{equation}
	   \boldsymbol{\mathcal{A}}^{\xi}(\mathbf{r},\mathbf{r'})=\int\frac{d^3k}{(2\pi)^3}e^{i\mathbf{k}\cdot(\mathbf{r}-\mathbf{r'})}\sum_{\lambda}\left( \sigma_0+\lambda\frac{\boldsymbol{\sigma\cdot}\mathbf{k}_{\parallel}}{|\mathbf{k}_\parallel|}\right) \mathcal{A}^{\lambda,\xi}(k). \label{eq:spectral_coordinates}
	\end{equation}
Notice that because $\mathbf{k}_{\parallel}=(k_x,k_y)$, when we perform the integration the spectral function takes the form $\boldsymbol{\mathcal{A}}^{\xi}(\mathbf{r},\mathbf{r'})=\delta(z-z')\boldsymbol{\mathcal{A}}^{\xi}(\mathbf{x},\mathbf{x'})$, which looks similar to the decoupled form of the GF in Eq.\eqref{eq:GF_R}.

%_____________________________
\section{Linear Response Theory}
\label{app:linear}

The tensor $K_{\alpha\beta}$ is defined using the retarded current-current correlator in Eq.\eqref{eq:corr_current}. Introducing in Eq.(\ref{eq:corr_current}) the complete and orthonormal basis $\left\lbrace\left| \Psi_{\lambda,\mathbf{k}}\right\rangle\right\rbrace$ of the total Hamiltonian, such that $\hat{H}^{\xi}\left| \Psi_{\lambda,\mathbf{k}}\right\rangle =\mathcal{E}^{\lambda,\xi}_{\mathbf{k}}\left| \Psi_{\lambda,\mathbf{k}}\right\rangle$ and $\hat{\rho}\left| \Psi_{\lambda,\mathbf{k}}\right\rangle=\rho\left( \mathcal{E}^{\lambda,\xi}_{\mathbf{k}}\right) \left| \Psi_{\lambda,\mathbf{k}}\right\rangle$, we obtain
	\begin{align}
	 &K^{\xi}_{\alpha\beta}(\mathbf{r},t;\mathbf{r'},t')
		=i\hbar^{-1}\theta(t-t')\nonumber\\
  &\quad\times\int \frac{d^3k}{(2\pi)^3}\int \frac{d^3k'}{(2\pi)^3}\sum_{\lambda,\lambda'} \left[ \rho\left(\mathcal{E}^{\lambda,\xi}_{\mathbf{k}}\right)  -\rho \left(\mathcal{E}^{\lambda',\xi}_{\mathbf{k'}} \right) \right] \nonumber \\
		& \quad \times\left\langle \Psi_{\lambda,\mathbf{k}}\middle|\hat{j}_{\alpha}(\mathbf{r})\middle|\Psi_{\lambda',\mathbf{k'}}\right\rangle \left\langle \Psi_{\lambda',\mathbf{k'}}\middle|\hat{j}_{\beta}(\mathbf{r'})\middle|\Psi_{\lambda,\mathbf{k}}\right\rangle e^{\frac{i}{\hbar}\left( \mathcal{E}^{\lambda,\xi}_{\mathbf{k}}-\mathcal{E}^{\lambda',\xi}_{\mathbf{k'}}\right) (t-t')}. \label{eq:correlator_1}   
	\end{align}
Using the Fourier representation of the Heaviside step function we obtain the correlator in the frequency domain
	\begin{align}
		&K^{\xi}_{\alpha\beta}(\mathbf{r},\mathbf{r'};\omega)\nonumber\\
  &= \int \frac{d^3k}{(2\pi)^3}\int \frac{d^3k'}{(2\pi)^3}\sum_{\lambda,\lambda'} \frac{\left[ \rho \left(\mathcal{E}^{\lambda',\xi}_{\mathbf{k'}} \right)-\rho\left(\mathcal{E}^{\lambda,\xi}_{\mathbf{k}}\right)  \right] }{\hbar\omega+ \mathcal{E}^{\lambda,\xi}_{\mathbf{k}}-\mathcal{E}^{\lambda',\xi}_{\mathbf{k'}}+i\eta^{+}}\nonumber\\
  &\times\left\langle \Psi_{\lambda,\mathbf{k}}\middle|\hat{j}_{\alpha}(\mathbf{r})\middle|\Psi_{\lambda',\mathbf{k'}}\right\rangle \left\langle \Psi_{\lambda',\mathbf{k'}}\middle|\hat{j}_{\beta}(\mathbf{r'})\middle|\Psi_{\lambda,\mathbf{k}}\right\rangle . 
	\end{align}
	 The electric current density operator for the Weyl equation is $\mathbf{\hat{j}}^{\xi}(\mathbf{r})=-e\xi v_F  \left| \mathbf{r}\right\rangle  \boldsymbol{\sigma}\left\langle \mathbf{r}\right|$. Then
	\begin{align}
		&K^{\xi}_{\alpha\beta}(\mathbf{r},\mathbf{r'};\omega)\nonumber\\
  &\quad=e^2v_{F}^2\int \frac{d^3k}{(2\pi)^3}\int \frac{d^3k'}{(2\pi)^3}\sum_{\lambda,\lambda'} \frac{\left[ \rho \left(\mathcal{E}^{\lambda',\xi}_{\mathbf{k'}} \right)-\rho\left(\mathcal{E}^{\lambda,\xi}_{\mathbf{k}}\right)  \right] }{\hbar\omega+ \mathcal{E}^{\lambda,\xi}_{\mathbf{k}}-\mathcal{E}^{\lambda',\xi}_{\mathbf{k'}}+i\eta^{+}}\nonumber\\
  &\quad \phantom{=} \times \Tr\left[ \sigma_{\alpha}\Psi_{\lambda',\mathbf{k'}}(\mathbf{r})\otimes\Psi_{\lambda',\mathbf{k'}}^{\dagger}(\mathbf{r'})\sigma_{\beta} \Psi_{\lambda,\mathbf{k}}(\mathbf{r'})\otimes\Psi_{\lambda,\mathbf{k}}^{\dagger}(\mathbf{r})\right].
	\end{align}
We can rewrite this last expression in terms of the spectral function and the retarded/advanced GFs as follows
	\begin{align}
		&K^{\xi}_{\alpha\beta}(\mathbf{r},\mathbf{r'};\omega)=-2e^2v_{F}^2\int_{-\infty}^{\infty} \frac{dE}{2\pi}f_{0}(E)\nonumber\\
  &\quad\times\Tr\left[ \sigma_{\alpha}\boldsymbol{\mathcal{A}}^{\xi}(\mathbf{r},\mathbf{r'};E)\sigma_{\beta}\mathbf{G}^{\xi}_{A }(\mathbf{r'},\mathbf{r};E-\hbar\omega)\right.\nonumber\\
  &\quad \quad \quad \quad \left.+\sigma_{\alpha}\mathbf{G}^{\xi}_{R}(\mathbf{r},\mathbf{r'};E+\hbar\omega)\sigma_{\beta}\boldsymbol{\mathcal{A}}^{\xi}(\mathbf{r'},\mathbf{r};E)\right],
	\end{align}
 where the additional factor of 2 is due to the spin degeneracy and $f_0(E)$ is the Fermi distribution. In the first term, we can shift energy variable $E\rightarrow E+\hbar\omega$, such that
	\begin{align}
		&K^{\xi}_{\alpha\beta}(\mathbf{r},\mathbf{r'};\omega)=-2e^2v_{F}^2\int_{-\infty}^{\infty} \frac{dE}{2\pi}\nonumber\\
  &\quad \times\left[ f_{0}(E+\hbar\omega)\Tr \sigma_{\alpha}\boldsymbol{\mathcal{A}}^{\xi}(\mathbf{r},\mathbf{r'};E+\hbar\omega)\sigma_{\beta}\mathbf{G}^{\xi}_{A }(\mathbf{r'},\mathbf{r};E)\right. \nonumber\\
		&\quad \quad \quad  \left. +f_{0}(E)\Tr \sigma_{\alpha}\mathbf{G}^{\xi}_{R}(\mathbf{r},\mathbf{r'};E+\hbar\omega)\sigma_{\beta}\boldsymbol{\mathcal{A}}^{\xi}(\mathbf{r'},\mathbf{r};E)\right]. \label{eq:corr_E+hw}
	\end{align}
We are interested in the real part of the conductivity tensor. Then
	\begin{align}
		&\Re\,\sigma^{\xi}_{\alpha\beta}(\mathbf{r},\mathbf{r'};\omega)=\Re\, \left( \frac{1}{i\omega} K^{\xi}_{\alpha\beta}(\mathbf{r},\mathbf{r'};\omega)\right) \nonumber\\
		&\quad =-\frac{i}{2\omega}\left[ K^{\xi}_{\alpha\beta}(\mathbf{r},\mathbf{r'};\omega)-K^{\xi \dagger}_{\beta\alpha}(\mathbf{r},\mathbf{r'};\omega)\right] ,
	\end{align}
	where we have used that $K^{*}_{\alpha \beta}=K^{\dagger}_{\beta\alpha }$. Then, taking the Hermitian conjugate of Eq.(\ref{eq:corr_E+hw}) and using the cyclic property of the trace we can write
	\begin{align}
		\Re\,	\sigma^{\xi}_{\alpha\beta}(\mathbf{r},\mathbf{r'};\omega)&=
		 -\frac{e^2\hbar v_{F}^2}{2\pi} \int_{-\infty}^{\infty} dE \left\lbrace \frac{f_{0}(E+\hbar\omega)-f_{0}(E)}{\hbar\omega}\right\rbrace\nonumber\\
   &\quad \times\Tr \sigma_{\alpha}\boldsymbol{\mathcal{A}}^{\xi}(\mathbf{r},\mathbf{r'};E+\hbar\omega)\sigma_{\beta}\boldsymbol{\mathcal{A}}^{\xi}(\mathbf{r'},\mathbf{r};E). \label{eq:sigma_coordinates}
	\end{align}
	Introducing the spectral density in Eq.\eqref{eq:spectral_coordinates} in the DC conductivity expression in Eq.\eqref{eq:sigma_coordinates} we have
	\begin{align}
		&\Re\,\sigma^{\xi}_{\alpha\beta}(\mathbf{r},\mathbf{r'};\omega)=\int \frac{d^3q}{(2\pi)^3}e^{i\mathbf{q}\cdot(\mathbf{r}-\mathbf{r'})}\left[-\frac{e^2\hbar v_{F}^2}{2\pi }\int \frac{d^3k'}{(2\pi)^3} \right.\nonumber\\
  &\quad\times\int_{-\infty}^{\infty} dE\,   \left\lbrace \frac{f_{0}(E+\hbar\omega)-f_{0}(E)}{\hbar\omega}\right\rbrace   \nonumber\\
       &\quad \times \sum_{\lambda,\lambda'}\Tr \left\lbrace \sigma_{\alpha}\left( \sigma_0+\lambda\frac{\boldsymbol{\sigma\cdot}(\mathbf{k}_{\parallel}+\mathbf{q})}{|\mathbf{k}_{\parallel}+\mathbf{q}|}\right)\sigma_{\beta}\left( \sigma_0+\lambda'\frac{\boldsymbol{\sigma\cdot}\mathbf{k}_{\parallel}}{|\mathbf{k}_{\parallel}|}\right) \right\rbrace\nonumber\\
    &\quad \left.\phantom{-\frac{e^2\hbar v_{F}^2}{2\pi }}\times \mathcal{A}^{\lambda,\xi}(|\mathbf{k}_{\parallel}+\mathbf{q}|;E+\hbar\omega)\mathcal{A}^{\lambda',\xi}(|\mathbf{k}_{\parallel}|;E)\right].
	\end{align}
 Then, the Fourier transform to the momentum space of the conductivity is the result in Eq.\eqref{eq:condcutivity_q}. We are computing the DC conductivity, so we take the limit $\mathbf{q}\rightarrow\mathbf{0}$ first and then the limit $\omega\rightarrow0$. The result is 
 \begin{align}
     &\sigma^{\xi}_{\alpha\beta}(T)=-\frac{e^2\hbar v_{F}^2}{2\pi }\int \frac{d^3k}{(2\pi)^3} \int_{-\infty}^{\infty} dE   \frac{\partial f_0(E)}{E} \nonumber\\
    &\quad \times \sum_{\lambda}\Tr \left\lbrace \sigma_{\alpha}\left( \sigma_0+\lambda\frac{\boldsymbol{\sigma\cdot}\mathbf{k}_{\parallel}}{|\mathbf{k}_{\parallel}|}\right)\sigma_{\beta}\left( \sigma_0+\lambda\frac{\boldsymbol{\sigma\cdot}\mathbf{k}_{\parallel}}{|\mathbf{k}_{\parallel}|}\right) \right\rbrace\nonumber\\
    &\quad \times \mathcal{A}^{\lambda,\xi}(|\mathbf{k}_{\parallel}|;E)\mathcal{A}^{\lambda,\xi}(|\mathbf{k}_{\parallel}|;E). 
 \end{align} 
where we have used Eq.\eqref{eq:A^2}. Now, we perform the trace
	\begin{align}
		&\Tr \left\lbrace \sigma_{\alpha}\left( \sigma_0+\lambda\frac{\boldsymbol{\sigma\cdot}\mathbf{k}_{\parallel}}{|\mathbf{k}_{\parallel}|}\right)\sigma_{\beta}\left( \sigma_0+\lambda\frac{\boldsymbol{\sigma\cdot}\mathbf{k}_{\parallel}}{|\mathbf{k}_{\parallel}|}\right) \right\rbrace\nonumber\\
        &\quad =\Tr\left\lbrace \sigma_{\alpha}\sigma_{\beta} +\lambda \left( \sigma_{\alpha}\sigma_{\beta}\sigma_{\gamma}+\sigma_{\alpha}\sigma_{\gamma}\sigma_{\beta}\right)\frac{k_{\gamma}}{|\mathbf{k}_{\parallel}|}\right.\nonumber\\
        &\quad \quad \quad \quad \left. +\lambda^2 \sigma_{\alpha}\sigma_{\gamma}\sigma_{\beta}\sigma_{\gamma'}\frac{k_{\gamma}k_{\gamma'}}{|\mathbf{k}_{\parallel}|^2}\right\rbrace  \nonumber \\
		&\quad =4\frac{k_{\alpha}k_{\beta}}{|\mathbf{k}_{\parallel}|^2},
	\end{align}
	where we have used the Pauli matrices trace technology  and that $\lambda^2=1$. The angular integration is performed immediately as follows
	\begin{equation}
		\int d\Omega  \,\, k_{\alpha}k_{\beta}=k^2\int_{0}^{\pi} d\theta \sin \theta \int_{0}^{2\pi} d\phi  \,\, n_{\alpha}n_{\beta}=2\pi k^2\delta_{\alpha\beta}, \label{eq:isotropy}
	\end{equation} 
	where $n_{\alpha}$ is the component of the unit vector $\mathbf{n}$ along the $\alpha$-direction (on the plane) and $\mathbf{k}_{\parallel}=k\mathbf{n}$. Then, we have
	\begin{align}
	    \sigma^{\xi}_{\alpha\beta}(T)&=-\delta_{\alpha\beta} \frac{4e^2\hbar v_{F}^2}{(2\pi)^3k_B T}\sum_{\lambda}\int_{-\infty}^{\infty}dE\,\, f_0(E)\left[ 1-f_0(E) \right] \notag \\
	    &\quad \times \int_{0}^{\infty}  dk \left[ 	\left\langle G_R^{\lambda,\xi}(\mathbf{k}_{\parallel})\right\rangle-	\left\langle G_A^{\lambda,\xi}(\mathbf{k}_{\parallel})\right\rangle\right]\nonumber\\
     & \quad \times\left[ 	\left\langle G_R^{\lambda,\xi}(\mathbf{k}_{\parallel})\right\rangle-	\left\langle G_A^{\lambda,\xi}(\mathbf{k}_{\parallel})\right\rangle\right]\mathbf{k}_{\parallel}\cdot\mathbf{k}_{\parallel} , \label{eq:conductivity_tensor_vertex}	
	    \end{align}
     where we have used the definition of the spectral function in terms of the retarded and advanced GFs. In the limit of low concentration of defects, i.e.,  $n_d\rightarrow 0$, we have that the unique leading contribution to the conductivity are given by the combination
	\begin{align}
	    &\left\langle G_R^{\lambda,\xi}(\mathbf{k}_{\parallel})\right\rangle\left\langle G_A^{\lambda,\xi}(\mathbf{k}_{\parallel})\right\rangle= \frac{1}{\left(E-\lambda\xi\hbar v_F k\right)^2+\left(\frac{\hbar}{2\tau^{(\lambda,\xi)}(k)}\right)^2}\notag \\
     &\quad \rightarrow\frac{2\pi \tau^{(\lambda,\xi)}(k)}{\hbar}\delta(E-\lambda\xi\hbar v_F k), \label{eq:limit_GRGA}
	\end{align}
	where we have used that the self-energy is purely imaginary and its relation with the relaxation time. The other contributions are negligible because they are not singular in $n_d$. For instance, in the low concentration limit, we have a contribution for two retarded GFs of the form
    \begin{align}
	    \left\langle G_R^{\lambda,\xi}(\mathbf{k}_{\parallel})\right\rangle\left\langle G_R^{\lambda,\xi}(\mathbf{k}_{\parallel})\right\rangle&= \frac{\left(E-\lambda\xi\hbar v_F k\right)^2-\left(\frac{\hbar}{2\tau^{(\lambda,\xi)}(k)}\right)^2}{\left\lbrace\left(E-\lambda\xi\hbar v_F k\right)^2+\left(\frac{\hbar}{2\tau^{(\lambda,\xi)}(k)}\right)^2\right\rbrace^2}\notag\\
     &+i\left(E-\lambda\xi\hbar v_F k\right)2\pi\delta(E-\lambda\xi\hbar v_F k). \label{eq:limit_GRGR}
	\end{align}
	The second term is zero and the first term is a function centered at $E-\lambda\xi\hbar v_F k$, but when integrated over $k$ it gives zero. Something similar occurs with the contribution of two advanced GFs. The result is the diagonal conductivity tensor given in Eq.\eqref{eq:conductivity_magnit_no_vertex}. 

%%%%%%%%%%%%%%%%%%%%%%%%%%%%%%%%%%%%%%%%%
\vspace{2mm}

\emph{Acknowledgments:} This research was funded by Fondecyt grant number 1190361 and by ANID PIA Anillo ACT/192023. 

%%%%%%%%%%%%%%%%%%%%%%%%%%%%%%%%%%%%%%%%%

\bibliographystyle{apsrev4-1} 
\bibliography{Dislocations}
	
\end{document}